# GW190425: Observation of a Compact Binary Coalescence with Total Mass $\sim 3.4 M_{\odot}$

B. P. Abbott[1], R. Abbott[1], T. D. Abbott[2], S. Abraham[3], F. Acernese[4,5], K. Ackley[6], C. Adams[7], R. X. Adhikari[1], V. B. Adya[8], C. Affeldt[9,10], M. Agathos[11,12], K. Agatsuma[13], N. Aggarwal[14], O. D. Aguiar[15], L. Aiello[16,17], A. Ain[3], P. Ajith[18], G. Allen[19], A. Allocca[20,21], M. A. Aloy[22], P. A. Altin[8], A. Amato[23], S. Anand[1], A. Ananyeva[1], S. B. Anderson[1], W. G. Anderson[24], S. V. Angelova[25], S. Antier[26], S. Appert[1], K. Arai[1], M. C. Araya[1], J. S. Areeda[27], M. Arène[26], N. Arnaud[28,29], S. M. Aronson[30], K. G. Arun[31], S. Ascenzi[16,32], G. Ashton[6], S. M. Aston[7], P. Astone[33], F. Aubin[34], P. Aufmuth[10], K. AultONeal[35], C. Austin[2], V. Avendano[36], A. Avila-Alvarez[27], S. Babak[26], P. Bacon[26], F. Badaracco[16,17], M. K. M. Bader[37], S. Bae[38], J. Baird[26], P. T. Baker[39], F. Baldaccini[40,41], G. Ballardin[29], S. W. Ballmer[42], A. Bals[35], S. Banagiri[43], J. C. Barayoga[1], C. Barbieri[44,45], S. E. Barclay[46], B. C. Barish[1], D. Barker[47], K. Barkett[48], S. Barnum[14], F. Barone[5,49], B. Barr[46], L. Barsotti[14], M. Barsuglia[26], D. Barta[50], J. Bartlett[47], I. Bartos[30], R. Bassiri[51], A. Basti[20,21], M. Bawaj[41,52], J. C. Bayley[46], A. C. Baylor[53], M. Bazzan[54,55], B. Bécsy[56], M. Bejger[26,57], I. Belahcene[28], A. S. Bell[46], D. Beniwal[58], M. G. Benjamin[35], B. K. Berger[51], G. Bergmann[9,10], S. Bernuzzi[11], C. P. L. Berry[59], D. Bersanetti[60], A. Bertolini[37], J. Betzwieser[7], R. Bhandare[61], J. Bidler[27], E. Biggs[24], I. A. Bilenko[62], S. A. Bilgili[39], G. Billingsley[1], R. Birney[25], O. Birnholtz[63], S. Biscans[1,14], M. Bischi[64,65], S. Biscoveanu[14], A. Bisht[10], M. Bitossi[21,29], M. A. Bizouard[66], J. K. Blackburn[1], J. Blackman[48], C. D. Blair[7], D. G. Blair[67], R. M. Blair[47], S. Bloemen[68], F. Bobba[69,70], N. Bode[9,10], M. Boer[66], Y. Boetzel[71], G. Bogaert[66], F. Bondu[72], R. Bonnand[34], P. Booker[9,10], B. A. Boom[37], R. Bork[1], V. Boschi[29], S. Bose[3], V. Bossilkov[67], J. Bosveld[67], Y. Bouffanais[54,55], A. Bozzi[29], C. Bradaschia[21], P. R. Brady[24], A. Bramley[7], M. Branchesi[16,17], J. E. Brau[73], M. Breschi[11], T. Briant[74], J. H. Briggs[46], F. Brighenti[64,65], A. Brillet[66], M. Brinkmann[9,10], P. Brockill[24], A. F. Brooks[1], J. Brooks[29], D. D. Brown[58], S. Brunett[1], A. Buikema[14], T. Bulik[75], H. J. Bulten[37,76], A. Buonanno[77,78], D. Buskulic[34], C. Buy[26], R. L. Byer[51], M. Cabero[9,10], L. Cadonati[79], G. Cagnoli[80], C. Cahillane[1], J. Calderón Bustillo[6], T. A. Callister[1], E. Calloni[5,81], J. B. Camp[82], W. A. Campbell[6], M. Canepa[60,83], K. C. Cannon[84], H. Cao[58], J. Cao[85], G. Carapella[69,70], F. Carbognani[29], S. Caride[86], M. F. Carney[59], G. Carullo[20,21], J. Casanueva Diaz[21], C. Casentini[32,87], S. Caudill[37], M. Cavaglià[88,89], F. Cavalier[28], R. Cavalieri[29], G. Cella[21], P. Cerdá-Durán[22], E. Cesarini[32,90], O. Chaibi[66], K. Chakravarti[3], S. J. Chamberlin[91], M. Chan[46], S. Chao[92], P. Charlton[93], E. A. Chase[59], E. Chassande-Mottin[26], D. Chatterjee[24], M. Chaturvedi[61], K. Chatziioannou[94], B. D. Cheeseboro[39], H. Y. Chen[95], X. Chen[67], Y. Chen[48], H.-P. Cheng[30], C. K. Cheong[96], H. Y. Chia[30], F. Chiadini[70,97], A. Chincarini[60], A. Chiummo[29], G. Cho[98], H. S. Cho[99], M. Cho[78], N. Christensen[66,100], Q. Chu[67], S. Chua[74], K. W. Chung[96], S. Chung[67], G. Ciani[54,55], M. Cieślar[57], A. A. Ciobanu[58], R. Ciolfi[55,101], F. Cipriano[66], A. Cirone[60,83], F. Clara[47], J. A. Clark[79], P. Clearwater[102], F. Cleva[66], E. Coccia[16,17], P.-F. Cohadon[74], D. Cohen[28], M. Colleoni[103], C. G. Collette[104], C. Collins[13], M. Colpi[44,45], L. R. Cominsky[105], M. Constancio, Jr.[15], L. Conti[55], S. J. Cooper[13], P. Corban[7], T. R. Corbitt[2], I. Cordero-Carrión[106], S. Corezzi[40,41], K. R. Corley[107], N. Cornish[56], D. Corre[28], A. Corsi[86], S. Cortese[29], C. A. Costa[15], R. Cotesta[77], M. W. Coughlin[1], S. B. Coughlin[59,108], J.-P. Coulon[66], S. T. Countryman[107], P. Couvares[1], P. B. Covas[103], E. E. Cowan[79], D. M. Coward[67], M. J. Cowart[7], D. C. Coyne[1], R. Coyne[109], J. D. E. Creighton[24], T. D. Creighton[110], J. Cripe[2], M. Croquette[74], S. G. Crowder[111], T. J. Cullen[2], A. Cumming[46], L. Cunningham[46], E. Cuoco[29], T. Dal Canton[82], G. Dálya[112], B. D'Angelo[60,83], S. L. Danilishin[9,10], S. D'Antonio[32], K. Danzmann[9,10], A. Dasgupta[113], C. F. Da Silva Costa[30], L. E. H. Datrier[46], V. Dattilo[29], I. Dave[61], M. Davier[28], D. Davis[42], E. J. Daw[114], D. DeBra[51], M. Deenadayalan[3], J. Degallaix[23], M. De Laurentis[5,81], S. Deléglise[74], N. De Lillo[46], W. Del Pozzo[20,21], L. M. DeMarchi[59], N. Demos[14], T. Dent[115], R. De Pietri[116,117], R. De Rosa[5,81], C. De Rossi[23,29], R. DeSalvo[118], O. de Varona[9,10], S. Dhurandhar[3], M. C. Díaz[110], T. Dietrich[37], L. Di Fiore[5], C. DiFronzo[13], C. Di Giorgio[69,70], F. Di Giovanni[22], M. Di Giovanni[119,120], T. Di Girolamo[5,81], A. Di Lieto[20,21], B. Ding[104], S. Di Pace[33,121], I. Di Palma[33,121], F. Di Renzo[20,21], A. K. Divakarla[30], A. Dmitriev[13], Z. Doctor[95], F. Donovan[14], K. L. Dooley[88,108], S. Doravari[3], I. Dorrington[108], T. P. Downes[24], M. Drago[16,17], J. C. Driggers[47], Z. Du[85], J.-G. Ducoin[28], R. Dudi[77], P. Dupej[46], O. Durante[69,70], S. E. Dwyer[47], P. J. Easter[6], G. Eddolls[46], T. B. Edo[114], A. Effler[7], P. Ehrens[1], J. Eichholz[8], S. S. Eikenberry[30], M. Eisenmann[34], R. A. Eisenstein[14], L. Errico[5,81], R. C. Essick[95], H. Estelles[103], D. Estevez[34], Z. B. Etienne[39], T. Etzel[1], M. Evans[14], T. M. Evans[7], V. Fafone[16,32,87], S. Fairhurst[108], X. Fan[85], S. Farinon[60], B. Farr[73], W. M. Farr[13], E. J. Fauchon-Jones[108], M. Favata[36], M. Fays[114], M. Fazio[122], C. Fee[123], J. Feicht[1], M. M. Fejer[51], F. Feng[26], A. Fernandez-Galiana[14], I. Ferrante[20,21], E. C. Ferreira[15], T. A. Ferreira[15], F. Fidecaro[20,21], I. Fiori[29], D. Fiorucci[16,17], M. Fishbach[95], R. P. Fisher[124], J. M. Fishner[14], R. Fittipaldi[70,125], M. Fitz-Axen[43], V. Fiumara[70,126], R. Flaminio[34,127], M. Fletcher[46], E. Floden[43], E. Flynn[27], H. Fong[84], J. A. Font[22,128], P. W. F. Forsyth[8], J.-D. Fournier[66], Francisco Hernandez Vivanco[6], S. Frasca[33,121], F. Frasconi[21], Z. Frei[112], A. Freise[13], R. Frey[73], V. Frey[28], P. Fritschel[14], V. V. Frolov[7], G. Fronzè[129], P. Fulda[30], M. Fyffe[7], H. A. Gabbard[46], B. U. Gadre[77], S. M. Gaebel[13], J. R. Gair[130], R. Gamba[11], L. Gammaitoni[40], S. G. Gaonkar[3], C. García-Quirós[103], F. Garufi[5,81], B. Gateley[47], S. Gaudio[35], G. Gaur[131], V. Gayathri[132], G. Gemme[60], E. Genin[29], A. Gennai[21], D. George[19], J. George[61], R. George[133], L. Gergely[134], S. Ghonge[79], Abhirup Ghosh[77], Archisman Ghosh[37], S. Ghosh[24], B. Giacomazzo[119,120], J. A. Giaime[2,7], K. D. Giardina[7], D. R. Gibson[135], K. Gill[107], L. Glover[136], J. Gniesmer[137], P. Godwin[91], E. Goetz[47], R. Goetz[30], B. Goncharov[6], G. González[2], J. M. Gonzalez Castro[20,21], A. Gopakumar[138], S. E. Gossan[1], M. Gosselin[20,21,29], R. Gouaty[34], B. Grace[8], A. Grado[5,139], M. Granata[23], A. Grant[46], S. Gras[14], P. Grassia[1], C. Gray[47], R. Gray[46], G. Greco[64,65], A. C. Green[30], R. Green[108], E. M. Gretarsson[35], A. Grimaldi[119,120], S. J. Grimm[16,17],

P. Groot[68], H. Grote[108], S. Grunewald[77], P. Gruning[28], G. M. Guidi[64,65], H. K. Gulati[113], Y. Guo[37], A. Gupta[91], Anchal Gupta[1], P. Gupta[37], E. K. Gustafson[1], R. Gustafson[140], L. Haegel[103], O. Halim[16,17], B. R. Hall[141], E. D. Hall[14], E. Z. Hamilton[108], G. Hammond[46], M. Haney[71], M. M. Hanke[9,10], J. Hanks[47], C. Hanna[91], M. D. Hannam[108], O. A. Hannuksela[96], T. J. Hansen[35], J. Hanson[7], T. Harder[66], T. Hardwick[2], K. Haris[18], J. Harms[16,17], G. M. Harry[142], I. W. Harry[143], R. K. Hasskew[7], C. J. Haster[14], K. Haughian[46], F. J. Hayes[46], J. Healy[63], A. Heidmann[74], M. C. Heintze[7], H. Heitmann[66], F. Hellman[144], P. Hello[28], G. Hemming[29], M. Hendry[46], I. S. Heng[46], J. Hennig[9,10], M. Heurs[9,10], S. Hild[46], T. Hinderer[37,145,146], W. C. G. Ho[147], S. Hochheim[9,10], D. Hofman[23], A. M. Holgado[19], N. A. Holland[8], K. Holt[7], D. E. Holz[95], P. Hopkins[108], C. Horst[24], J. Hough[46], E. J. Howell[67], C. G. Hoy[108], Y. Huang[14], M. T. Hübner[6], E. A. Huerta[19], D. Huet[28], B. Hughey[35], V. Hui[34], S. Husa[103], S. H. Huttner[46], T. Huynh-Dinh[7], B. Idzkowski[75], A. Iess[32,87], H. Inchauspe[30], C. Ingram[58], R. Inta[86], G. Intini[33,121], B. Irwin[123], H. N. Isa[46], J.-M. Isac[74], M. Isi[14], B. R. Iyer[18], T. Jacqmin[74], S. J. Jadhav[148], K. Jani[79], N. N. Janthalur[148], P. Jaranowski[149], D. Jariwala[30], A. C. Jenkins[150], J. Jiang[30], D. S. Johnson[19], N. K. Johnson-McDaniel[12], A. W. Jones[13], D. I. Jones[151], J. D. Jones[47], R. Jones[46], R. J. G. Jonker[37], L. Ju[67], J. Junker[9,10], C. V. Kalaghatgi[108], V. Kalogera[59], B. Kamai[1], S. Kandhasamy[3], G. Kang[38], J. B. Kanner[1], S. J. Kapadia[24], S. Karki[73], R. Kashyap[18], M. Kasprzack[1], W. Kastaun[9,10], S. Katsanevas[29], E. Katsavounidis[14], W. Katzman[7], S. Kaufer[10], K. Kawabe[47], N. V. Keerthana[3], F. Kéfélian[66], D. Keitel[143], R. Kennedy[114], J. S. Key[152], F. Y. Khalili[62], I. Khan[16,32], S. Khan[9,10], E. A. Khazanov[153], N. Khetan[16,17], M. Khursheed[61], N. Kijbunchoo[8], Chunglee Kim[154], J. C. Kim[155], K. Kim[96], W. Kim[58], W. S. Kim[156], Y.-M. Kim[157], C. Kimball[59], P. J. King[47], M. Kinley-Hanlon[46], R. Kirchhoff[9,10], J. S. Kissel[47], L. Kleybolte[137], J. H. Klika[24], S. Klimenko[30], T. D. Knowles[39], P. Koch[9,10], S. M. Koehlenbeck[9,10], G. Koekoek[37,158], S. Koley[37], V. Kondrashov[1], A. Kontos[159], N. Koper[9,10], M. Korobko[137], W. Z. Korth[1], M. Kovalam[67], D. B. Kozak[1], C. Krämer[9,10], V. Kringel[9,10], N. Krishnendu[31], A. Królak[160,161], N. Krupinski[24], G. Kuehn[9,10], A. Kumar[148], P. Kumar[162], Rahul Kumar[47], Rakesh Kumar[113], L. Kuo[92], A. Kutynia[160], S. Kwang[24], B. D. Lackey[77], D. Laghi[20,21], K. H. Lai[96], T. L. Lam[96], M. Landry[47], P. Landry[27], B. B. Lane[14], R. N. Lang[163], J. Lange[63], B. Lantz[51], R. K. Lanza[14], A. Lartaux-Vollard[28], P. D. Lasky[6], M. Laxen[7], A. Lazzarini[1], C. Lazzaro[55], P. Leaci[33,121], S. Leavey[9,10], Y. K. Lecoeuche[47], C. H. Lee[99], H. K. Lee[164], H. M. Lee[165], H. W. Lee[155], J. Lee[98], K. Lee[46], J. Lehmann[9,10], A. K. Lenon[39], N. Leroy[28], N. Letendre[34], Y. Levin[6], A. Li[96], J. Li[85], K. J. L. Li[96], T. G. F. Li[96], X. Li[48], F. Lin[6], F. Linde[37,166], S. D. Linker[136], T. B. Littenberg[167], J. Liu[67], X. Liu[24], M. Llorens-Monteagudo[22], R. K. L. Lo[1,96], L. T. London[14], A. Longo[168,169], M. Lorenzini[16,17], V. Loriette[170], M. Lormand[7], G. Losurdo[21], J. D. Lough[9,10], C. O. Lousto[63], G. Lovelace[27], M. E. Lower[171], J. F. Lucaccioni[123], H. Lück[9,10], D. Lumaca[32,87], A. P. Lundgren[143], R. Lynch[14], Y. Ma[48], R. Macas[108], S. Macfoy[25], M. MacInnis[14], D. M. Macleod[108], A. Macquet[66], I. Magaña Hernandez[24], F. Magaña-Sandoval[30], R. M. Magee[91], E. Majorana[33], I. Maksimovic[170], A. Malik[61], N. Man[66], V. Mandic[43], V. Mangano[33,46,121], G. L. Mansell[14,47], M. Manske[24], M. Mantovani[29], M. Mapelli[54,55], F. Marchesoni[41,52], F. Marion[34], S. Márka[107], Z. Márka[107], C. Markakis[19], A. S. Markosyan[51], A. Markowitz[1], E. Maros[1], A. Marquina[106], S. Marsat[26], F. Martelli[64,65], I. W. Martin[46], R. M. Martin[36], V. Martinez[80], D. V. Martynov[13], H. Masalehdan[137], K. Mason[14], E. Massera[114], A. Masserot[34], T. J. Massinger[1], M. Masso-Reid[46], S. Mastrogiovanni[26], A. Matas[77], F. Matichard[1,14], L. Matone[107], N. Mavalvala[14], J. J. McCann[67], R. McCarthy[47], D. E. McClelland[8], S. McCormick[7], L. McCuller[14], S. C. McGuire[172], C. McIsaac[143], J. McIver[1], D. J. McManus[8], T. McRae[8], S. T. McWilliams[39], D. Meacher[24], G. D. Meadors[6], M. Mehmet[9,10], A. K. Mehta[18], J. Meidam[37], E. Mejuto Villa[70,118], A. Melatos[102], G. Mendell[47], R. A. Mercer[24], L. Mereni[23], K. Merfeld[73], E. L. Merilh[47], M. Merzougui[66], S. Meshkov[1], C. Messenger[46], C. Messick[91], F. Messina[44,45], R. Metzdorff[74], P. M. Meyers[102], F. Meylahn[9,10], A. Miani[119,120], H. Miao[13], C. Michel[23], H. Middleton[102], L. Milano[5,81], A. L. Miller[30,33,121], M. Millhouse[102], J. C. Mills[108], M. C. Milovich-Goff[136], O. Minazzoli[66,173], Y. Minenkov[32], A. Mishkin[30], C. Mishra[174], T. Mistry[114], S. Mitra[3], V. P. Mitrofanov[62], G. Mitselmakher[30], R. Mittleman[14], G. Mo[100], D. Moffa[123], K. Mogushi[88], S. R. P. Mohapatra[14], M. Molina-Ruiz[144], M. Mondin[136], M. Montani[64,65], C. J. Moore[13], D. Moraru[47], F. Morawski[57], G. Moreno[47], S. Morisaki[84], B. Mours[34], C. M. Mow-Lowry[13], F. Muciaccia[33,121], Arunava Mukherjee[9,10], D. Mukherjee[24], S. Mukherjee[110], Subroto Mukherjee[113], N. Mukund[3,9,10], A. Mullavey[7], J. Munch[58], E. A. Muñiz[42], M. Muratore[35], P. G. Murray[46], A. Nagar[90,129,175], I. Nardecchia[32,87], L. Naticchioni[33,121], R. K. Nayak[176], B. F. Neil[67], J. Neilson[70,118], G. Nelemans[37,68], T. J. N. Nelson[7], M. Nery[9,10], A. Neunzert[140], L. Nevin[1], K. Y. Ng[14], S. Ng[58], C. Nguyen[26], P. Nguyen[73], D. Nichols[37,145], S. A. Nichols[2], S. Nissanke[37,145], F. Nocera[29], C. North[108], L. K. Nuttall[143], M. Obergaulinger[22,177], J. Oberling[47], B. D. O'Brien[30], G. Oganesyan[16,17], G. H. Ogin[178], J. J. Oh[156], S. H. Oh[156], F. Ohme[9,10], H. Ohta[84], M. A. Okada[15], M. Oliver[103], P. Oppermann[9,10], Richard J. Oram[7], B. O'Reilly[7], R. G. Ormiston[43], L. F. Ortega[30], R. O'Shaughnessy[63], S. Ossokine[77], D. J. Ottaway[58], H. Overmier[7], B. J. Owen[86], A. E. Pace[91], G. Pagano[20,21], M. A. Page[67], G. Pagliaroli[16,17], A. Pai[132], S. A. Pai[61], J. R. Palamos[73], O. Palashov[153], C. Palomba[33], H. Pan[92], P. K. Panda[148], P. T. H. Pang[37,96], C. Pankow[59], F. Pannarale[33,121], B. C. Pant[61], F. Paoletti[21], A. Paoli[29], A. Parida[3], W. Parker[7,172], D. Pascucci[37,46], A. Pasqualetti[29], R. Passaquieti[20,21], D. Passuello[21], M. Patil[161], B. Patricelli[20,21], E. Payne[6], B. L. Pearlstone[46], T. C. Pechsiri[30], A. J. Pedersen[42], M. Pedraza[1], R. Pedurand[23,179], A. Pele[7], S. Penn[180], A. Perego[119,120], C. J. Perez[47], C. Périgois[34], A. Perreca[119,120], J. Petermann[137], H. P. Pfeiffer[77], M. Phelps[9,10], K. S. Phukon[3], O. J. Piccinni[33,121], M. Pichot[66], F. Piergiovanni[64,65], V. Pierro[70,118], G. Pillant[29], L. Pinard[23], I. M. Pinto[70,90,118], M. Pirello[47], M. Pitkin[181], W. Plastino[168,169], R. Poggiani[20,21], D. Y. T. Pong[96], S. Ponrathnam[3], P. Popolizio[29], E. K. Porter[26], J. Powell[171], A. K. Prajapati[113], J. Prasad[3], K. Prasai[51], R. Prasanna[148], G. Pratten[103], T. Prestegard[24], M. Principe[70,90,118], G. A. Prodi[119,120], L. Prokhorov[13], M. Punturo[41], P. Puppo[33], M. Pürrer[77], H. Qi[108], V. Quetschke[110], P. J. Quinonez[35], F. J. Raab[47], G. Raaijmakers[37,145], H. Radkins[47], N. Radulesco[66], P. Raffai[112], S. Raja[61], C. Rajan[61], B. Rajbhandari[86], M. Rakhmanov[110], K. E. Ramirez[110], A. Ramos-Buades[103], Javed Rana[3],

K. Rao[59], P. Rapagnani[33,121], V. Raymond[108], M. Razzano[20,21], J. Read[27], T. Regimbau[34], L. Rei[60], S. Reid[25], D. H. Reitze[1,30], P. Rettegno[129,182], F. Ricci[33,121], C. J. Richardson[35], J. W. Richardson[1], P. M. Ricker[19], G. Riemenschneider[129,182], K. Riles[140], M. Rizzo[59], N. A. Robertson[1,46], F. Robinet[28], A. Rocchi[32], L. Rolland[34], J. G. Rollins[1], V. J. Roma[73], M. Romanelli[72], R. Romano[4,5], C. L. Romel[47], J. H. Romie[7], C. A. Rose[24], D. Rose[27], K. Rose[123], M. J. B. Rosell[133], D. Rosińska[75], S. G. Rosofsky[19], M. P. Ross[183], S. Rowan[46], S. Roy[184], A. Rüdiger[9,10,198], P. Ruggi[29], G. Rutins[135], K. Ryan[47], S. Sachdev[91], T. Sadecki[47], M. Sakellariadou[150], O. S. Salafia[44,45,185], L. Salconi[29], M. Saleem[31], A. Samajdar[37], L. Sammut[6], E. J. Sanchez[1], L. E. Sanchez[1], N. Sanchis-Gual[186], J. R. Sanders[187], K. A. Santiago[36], E. Santos[66], N. Sarin[6], B. Sassolas[23], B. S. Sathyaprakash[91,108], O. Sauter[34,140], R. L. Savage[47], P. Schale[73], M. Scheel[48], J. Scheuer[59], P. Schmidt[13,68], R. Schnabel[137], R. M. S. Schofield[73], A. Schönbeck[137], E. Schreiber[9,10], B. W. Schulte[9,10], B. F. Schutz[108], J. Scott[46], S. M. Scott[8], E. Seidel[19], D. Sellers[7], A. S. Sengupta[184], N. Sennett[77], D. Sentenac[29], V. Sequino[60], A. Sergeev[153], Y. Setyawati[9,10], D. A. Shaddock[8], T. Shaffer[47], M. S. Shahriar[59], M. B. Shaner[136], A. Sharma[16,17], P. Sharma[61], P. Shawhan[78], H. Shen[19], R. Shink[188], D. H. Shoemaker[14], D. M. Shoemaker[79], K. Shukla[144], S. ShyamSundar[61], K. Siellez[79], M. Sieniawska[57], D. Sigg[47], L. P. Singer[82], D. Singh[91], N. Singh[75], A. Singhal[16,33], A. M. Sintes[103], S. Sitmukhambetov[110], V. Skliris[108], B. J. J. Slagmolen[8], T. J. Slaven-Blair[67], J. R. Smith[27], R. J. E. Smith[6], S. Somala[189], E. J. Son[156], S. Soni[2], B. Sorazu[46], F. Sorrentino[60], T. Souradeep[3], E. Sowell[86], A. P. Spencer[46], M. Spera[54,55], A. K. Srivastava[113], V. Srivastava[42], K. Staats[59], C. Stachie[66], M. Standke[9,10], D. A. Steer[26], M. Steinke[9,10], J. Steinlechner[46,137], S. Steinlechner[137], D. Steinmeyer[9,10], S. P. Stevenson[171], D. Stocks[51], R. Stone[110], D. J. Stops[13], K. A. Strain[46], G. Stratta[65,190], S. E. Strigin[62], A. Strunk[47], R. Sturani[191], A. L. Stuver[192], V. Sudhir[14], T. Z. Summerscales[193], L. Sun[1], S. Sunil[113], A. Sur[57], J. Suresh[84], P. J. Sutton[108], B. L. Swinkels[37], M. J. Szczepańczyk[35], M. Tacca[37], S. C. Tait[46], C. Talbot[6], D. B. Tanner[30], D. Tao[1], M. Tápai[134], A. Tapia[27], J. D. Tasson[100], R. Taylor[1], R. Tenorio[103], L. Terkowski[137], M. Thomas[7], P. Thomas[47], S. R. Thondapu[61], K. A. Thorne[7], E. Thrane[6], Shubhanshu Tiwari[119,120], Srishti Tiwari[138], V. Tiwari[108], K. Toland[46], M. Tonelli[20,21], Z. Tornasi[46], A. Torres-Forné[194], C. I. Torrie[1], D. Töyrä[13], F. Travasso[29,41], G. Traylor[7], M. C. Tringali[75], A. Tripathee[140], A. Trovato[26], L. Trozzo[21,195], K. W. Tsang[37], M. Tse[14], R. Tso[48], L. Tsukada[84], D. Tsuna[84], T. Tsutsui[84], D. Tuyenbayev[110], K. Ueno[84], D. Ugolini[196], C. S. Unnikrishnan[138], A. L. Urban[2], S. A. Usman[95], H. Vahlbruch[10], G. Vajente[1], G. Valdes[2], M. Valentini[119,120], N. van Bakel[37], M. van Beuzekom[37], J. F. J. van den Brand[37,76], C. Van Den Broeck[37,197], D. C. Vander-Hyde[42], L. van der Schaaf[37], J. V. VanHeijningen[67], A. A. van Veggel[46], M. Vardaro[54,55], V. Varma[48], S. Vass[1], M. Vasúth[50], A. Vecchio[13], G. Vedovato[55], J. Veitch[46], P. J. Veitch[58], K. Venkateswara[183], G. Venugopalan[1], D. Verkindt[34], F. Vetrano[64,65], A. Viceré[64,65], A. D. Viets[24], S. Vinciguerra[13], D. J. Vine[135], J.-Y. Vinet[66], S. Vitale[14], T. Vo[42], H. Vocca[40,41], C. Vorvick[47], S. P. Vyatchanin[62], A. R. Wade[1], L. E. Wade[123], M. Wade[123], R. Walet[37], M. Walker[27], L. Wallace[1], S. Walsh[24], H. Wang[13], J. Z. Wang[140], S. Wang[19], W. H. Wang[110], R. L. Ward[8], Z. A. Warden[35], J. Warner[47], M. Was[34], J. Watchi[104], B. Weaver[47], L.-W. Wei[9,10], M. Weinert[9,10], A. J. Weinstein[1], R. Weiss[14], F. Wellmann[9,10], L. Wen[67], E. K. Wessel[19], P. Weßels[9,10], J. W. Westhouse[35], K. Wette[8], J. T. Whelan[63], D. D. White[27], B. F. Whiting[30], C. Whittle[14], D. M. Wilken[9,10], D. Williams[46], A. R. Williamson[37,145], J. L. Willis[1], B. Willke[9,10], W. Winkler[9,10], C. C. Wipf[1], H. Wittel[9,10], G. Woan[46], J. Woehler[9,10], J. K. Wofford[63], J. L. Wright[46], D. S. Wu[9,10], D. M. Wysocki[63], S. Xiao[1], R. Xu[111], H. Yamamoto[1], C. C. Yancey[78], L. Yang[122], Y. Yang[30], Z. Yang[43], M. J. Yap[8], M. Yazback[30], D. W. Yeeles[108], Hang Yu[14], Haocun Yu[14], S. H. R. Yuen[96], A. K. Zadrożny[110], A. Zadrożny[160], M. Zanolin[35], T. Zelenova[29], J.-P. Zendri[55], M. Zevin[59], J. Zhang[67], L. Zhang[1], T. Zhang[46], C. Zhao[67], G. Zhao[104], M. Zhou[59], Z. Zhou[59], X. J. Zhu[6], A. B. Zimmerman[133], M. E. Zucker[1,14], and J. Zweizig[1,199]

[1] LIGO, California Institute of Technology, Pasadena, CA 91125, USA
[2] Louisiana State University, Baton Rouge, LA 70803, USA
[3] Inter-University Centre for Astronomy and Astrophysics, Pune 411007, India
[4] Dipartimento di Farmacia, Università di Salerno, I-84084 Fisciano, Salerno, Italy
[5] INFN, Sezione di Napoli, Complesso Universitario di Monte S.Angelo, I-80126 Napoli, Italy
[6] OzGrav, School of Physics & Astronomy, Monash University, Clayton 3800, Victoria, Australia
[7] LIGO Livingston Observatory, Livingston, LA 70754, USA
[8] OzGrav, Australian National University, Canberra, Australian Capital Territory 0200, Australia
[9] Max Planck Institute for Gravitational Physics (Albert Einstein Institute), D-30167 Hannover, Germany
[10] Leibniz Universität Hannover, D-30167 Hannover, Germany
[11] Theoretisch-Physikalisches Institut, Friedrich-Schiller-Universität Jena, D-07743 Jena, Germany
[12] University of Cambridge, Cambridge CB2 1TN, UK
[13] University of Birmingham, Birmingham B15 2TT, UK
[14] LIGO, Massachusetts Institute of Technology, Cambridge, MA 02139, USA
[15] Instituto Nacional de Pesquisas Espaciais, 12227-010 São José dos Campos, São Paulo, Brazil
[16] Gran Sasso Science Institute (GSSI), I-67100 L'Aquila, Italy
[17] INFN, Laboratori Nazionali del Gran Sasso, I-67100 Assergi, Italy
[18] International Centre for Theoretical Sciences, Tata Institute of Fundamental Research, Bengaluru 560089, India
[19] NCSA, University of Illinois at Urbana-Champaign, Urbana, IL 61801, USA
[20] Università di Pisa, I-56127 Pisa, Italy
[21] INFN, Sezione di Pisa, I-56127 Pisa, Italy
[22] Departamento de Astronomía y Astrofísica, Universitat de València, E-46100 Burjassot, València, Spain
[23] Laboratoire des Matériaux Avancés (LMA), CNRS/IN2P3, F-69622 Villeurbanne, France
[24] University of Wisconsin-Milwaukee, Milwaukee, WI 53201, USA
[25] SUPA, University of Strathclyde, Glasgow G1 1XQ, UK
[26] APC, AstroParticule et Cosmologie, Université Paris Diderot, CNRS/IN2P3, CEA/Irfu, Observatoire de Paris, Sorbonne Paris Cité, F-75205 Paris Cedex 13, France

[27] California State University Fullerton, Fullerton, CA 92831, USA
[28] LAL, Univ. Paris-Sud, CNRS/IN2P3, Université Paris-Saclay, F-91898 Orsay, France
[29] European Gravitational Observatory (EGO), I-56021 Cascina, Pisa, Italy
[30] University of Florida, Gainesville, FL 32611, USA
[31] Chennai Mathematical Institute, Chennai 603103, India
[32] INFN, Sezione di Roma Tor Vergata, I-00133 Roma, Italy
[33] INFN, Sezione di Roma, I-00185 Roma, Italy
[34] Laboratoire d'Annecy de Physique des Particules (LAPP), Univ. Grenoble Alpes, Université Savoie Mont Blanc, CNRS/IN2P3, F-74941 Annecy, France
[35] Embry-Riddle Aeronautical University, Prescott, AZ 86301, USA
[36] Montclair State University, Montclair, NJ 07043, USA
[37] Nikhef, Science Park 105, 1098 XG Amsterdam, The Netherlands
[38] Korea Institute of Science and Technology Information, Daejeon 34141, Republic of Korea
[39] West Virginia University, Morgantown, WV 26506, USA
[40] Università di Perugia, I-06123 Perugia, Italy
[41] INFN, Sezione di Perugia, I-06123 Perugia, Italy
[42] Syracuse University, Syracuse, NY 13244, USA
[43] University of Minnesota, Minneapolis, MN 55455, USA
[44] Università degli Studi di Milano-Bicocca, I-20126 Milano, Italy
[45] INFN, Sezione di Milano-Bicocca, I-20126 Milano, Italy
[46] SUPA, University of Glasgow, Glasgow G12 8QQ, UK
[47] LIGO Hanford Observatory, Richland, WA 99352, USA
[48] Caltech CaRT, Pasadena, CA 91125, USA
[49] Dipartimento di Medicina, Chirurgia e Odontoiatria "Scuola Medica Salernitana," Università di Salerno, I-84081 Baronissi, Salerno, Italy
[50] Wigner RCP, RMKI, H-1121 Budapest, Konkoly Thege Miklós út 29-33, Hungary
[51] Stanford University, Stanford, CA 94305, USA
[52] Università di Camerino, Dipartimento di Fisica, I-62032 Camerino, Italy
[53] Department of Physics and Astronomy, Dickinson College, Carlisle, PA 17013, USA
[54] Università di Padova, Dipartimento di Fisica e Astronomia, I-35131 Padova, Italy
[55] INFN, Sezione di Padova, I-35131 Padova, Italy
[56] Montana State University, Bozeman, MT 59717, USA
[57] Nicolaus Copernicus Astronomical Center, Polish Academy of Sciences, 00-716, Warsaw, Poland
[58] OzGrav, University of Adelaide, Adelaide, South Australia 5005, Australia
[59] Center for Interdisciplinary Exploration & Research in Astrophysics (CIERA), Northwestern University, Evanston, IL 60208, USA
[60] INFN, Sezione di Genova, I-16146 Genova, Italy
[61] RRCAT, Indore, Madhya Pradesh 452013, India
[62] Faculty of Physics, Lomonosov Moscow State University, Moscow 119991, Russia
[63] Rochester Institute of Technology, Rochester, NY 14623, USA
[64] Università degli Studi di Urbino "Carlo Bo," I-61029 Urbino, Italy
[65] INFN, Sezione di Firenze, I-50019 Sesto Fiorentino, Firenze, Italy
[66] Artemis, Université Côte d'Azur, Observatoire Côte d'Azur, CNRS, CS 34229, F-06304 Nice Cedex 4, France
[67] OzGrav, University of Western Australia, Crawley, Western Australia 6009, Australia
[68] Department of Astrophysics/IMAPP, Radboud University Nijmegen, P.O. Box 9010, 6500 GL Nijmegen, The Netherlands
[69] Dipartimento di Fisica "E.R. Caianiello," Università di Salerno, I-84084 Fisciano, Salerno, Italy
[70] INFN, Sezione di Napoli, Gruppo Collegato di Salerno, Complesso Universitario di Monte S. Angelo, I-80126 Napoli, Italy
[71] Physik-Institut, University of Zurich, Winterthurerstrasse 190, 8057 Zurich, Switzerland
[72] Univ Rennes, CNRS, Institut FOTON—UMR6082, F-3500 Rennes, France
[73] University of Oregon, Eugene, OR 97403, USA
[74] Laboratoire Kastler Brossel, Sorbonne Université, CNRS, ENS-Université PSL, Collège de France, F-75005 Paris, France
[75] Astronomical Observatory Warsaw University, 00-478 Warsaw, Poland
[76] VU University Amsterdam, 1081 HV Amsterdam, The Netherlands
[77] Max Planck Institute for Gravitational Physics (Albert Einstein Institute), D-14476 Potsdam-Golm, Germany
[78] University of Maryland, College Park, MD 20742, USA
[79] School of Physics, Georgia Institute of Technology, Atlanta, GA 30332, USA
[80] Université de Lyon, Université Claude Bernard Lyon 1, CNRS, Institut Lumière Matière, F-69622 Villeurbanne, France
[81] Università di Napoli "Federico II," Complesso Universitario di Monte S.Angelo, I-80126 Napoli, Italy
[82] NASA Goddard Space Flight Center, Greenbelt, MD 20771, USA
[83] Dipartimento di Fisica, Università degli Studi di Genova, I-16146 Genova, Italy
[84] RESCEU, University of Tokyo, Tokyo, 113-0033, Japan
[85] Tsinghua University, Beijing 100084, People's Republic of China
[86] Texas Tech University, Lubbock, TX 79409, USA
[87] Università di Roma Tor Vergata, I-00133 Roma, Italy
[88] The University of Mississippi, University, MS 38677, USA
[89] Missouri University of Science and Technology, Rolla, MO 65409, USA
[90] Museo Storico della Fisica e Centro Studi e Ricerche "Enrico Fermi," I-00184 Roma, Italy
[91] The Pennsylvania State University, University Park, PA 16802, USA
[92] National Tsing Hua University, Hsinchu City, 30013 Taiwan, People's Republic of China
[93] Charles Sturt University, Wagga Wagga, New South Wales 2678, Australia
[94] Canadian Institute for Theoretical Astrophysics, University of Toronto, Toronto, Ontario M5S 3H8, Canada
[95] University of Chicago, Chicago, IL 60637, USA
[96] The Chinese University of Hong Kong, Shatin, NT, Hong Kong
[97] Dipartimento di Ingegneria Industriale (DIIN), Università di Salerno, I-84084 Fisciano, Salerno, Italy
[98] Seoul National University, Seoul 08826, Republic of Korea
[99] Pusan National University, Busan 46241, Republic of Korea
[100] Carleton College, Northfield, MN 55057, USA
[101] INAF, Osservatorio Astronomico di Padova, I-35122 Padova, Italy
[102] OzGrav, University of Melbourne, Parkville, Victoria 3010, Australia

[103] Universitat de les Illes Balears, IAC3—IEEC, E-07122 Palma de Mallorca, Spain
[104] Université Libre de Bruxelles, Brussels 1050, Belgium
[105] Sonoma State University, Rohnert Park, CA 94928, USA
[106] Departamento de Matemáticas, Universitat de València, E-46100 Burjassot, València, Spain
[107] Columbia University, New York, NY 10027, USA
[108] Cardiff University, Cardiff CF24 3AA, UK
[109] University of Rhode Island, Kingston, RI 02881, USA
[110] The University of Texas Rio Grande Valley, Brownsville, TX 78520, USA
[111] Bellevue College, Bellevue, WA 98007, USA
[112] MTA-ELTE Astrophysics Research Group, Institute of Physics, Eötvös University, Budapest 1117, Hungary
[113] Institute for Plasma Research, Bhat, Gandhinagar 382428, India
[114] The University of Sheffield, Sheffield S10 2TN, UK
[115] IGFAE, Campus Sur, Universidade de Santiago de Compostela, E-15782, Spain
[116] Dipartimento di Scienze Matematiche, Fisiche e Informatiche, Università di Parma, I-43124 Parma, Italy
[117] INFN, Sezione di Milano Bicocca, Gruppo Collegato di Parma, I-43124 Parma, Italy
[118] Dipartimento di Ingegneria, Università del Sannio, I-82100 Benevento, Italy
[119] Università di Trento, Dipartimento di Fisica, I-38123 Povo, Trento, Italy
[120] INFN, Trento Institute for Fundamental Physics and Applications, I-38123 Povo, Trento, Italy
[121] Università di Roma "La Sapienza," I-00185 Roma, Italy
[122] Colorado State University, Fort Collins, CO 80523, USA
[123] Kenyon College, Gambier, OH 43022, USA
[124] Christopher Newport University, Newport News, VA 23606, USA
[125] CNR-SPIN, c/o Università di Salerno, I-84084 Fisciano, Salerno, Italy
[126] Scuola di Ingegneria, Università della Basilicata, I-85100 Potenza, Italy
[127] National Astronomical Observatory of Japan, 2-21-1 Osawa, Mitaka, Tokyo 181-8588, Japan
[128] Observatori Astronòmic, Universitat de València, E-46980 Paterna, València, Spain
[129] INFN Sezione di Torino, I-10125 Torino, Italy
[130] School of Mathematics, University of Edinburgh, Edinburgh EH9 3FD, UK
[131] Institute Of Advanced Research, Gandhinagar 382426, India
[132] Indian Institute of Technology Bombay, Powai, Mumbai 400 076, India
[133] Department of Physics, University of Texas, Austin, TX 78712, USA
[134] University of Szeged, Dóm tér 9, Szeged 6720, Hungary
[135] SUPA, University of the West of Scotland, Paisley PA1 2BE, UK
[136] California State University, Los Angeles, 5151 State University Dr, Los Angeles, CA 90032, USA
[137] Universität Hamburg, D-22761 Hamburg, Germany
[138] Tata Institute of Fundamental Research, Mumbai 400005, India
[139] INAF, Osservatorio Astronomico di Capodimonte, I-80131 Napoli, Italy
[140] University of Michigan, Ann Arbor, MI 48109, USA
[141] Washington State University, Pullman, WA 99164, USA
[142] American University, Washington, D.C. 20016, USA
[143] University of Portsmouth, Portsmouth, PO1 3FX, UK
[144] University of California, Berkeley, CA 94720, USA
[145] GRAPPA, Anton Pannekoek Institute for Astronomy and Institute for High-Energy Physics, University of Amsterdam, Science Park 904, 1098 XH Amsterdam, The Netherlands
[146] Delta Institute for Theoretical Physics, Science Park 904, 1090 GL Amsterdam, The Netherlands
[147] Department of Physics and Astronomy, Haverford College, 370 Lancaster Avenue, Haverford, PA 19041, USA
[148] Directorate of Construction, Services & Estate Management, Mumbai 400094, India
[149] University of Białystok, 15-424 Białystok, Poland
[150] King's College London, University of London, London WC2R 2LS, UK
[151] University of Southampton, Southampton SO17 1BJ, UK
[152] University of Washington Bothell, Bothell, WA 98011, USA
[153] Institute of Applied Physics, Nizhny Novgorod, 603950, Russia
[154] Ewha Womans University, Seoul 03760, Republic of Korea
[155] Inje University Gimhae, South Gyeongsang 50834, Republic of Korea
[156] National Institute for Mathematical Sciences, Daejeon 34047, Republic of Korea
[157] Ulsan National Institute of Science and Technology, Ulsan 44919, Republic of Korea
[158] Maastricht University, P.O. Box 616, 6200 MD Maastricht, The Netherlands
[159] Bard College, 30 Campus Rd, Annandale-On-Hudson, NY 12504, USA
[160] NCBJ, 05-400 Świerk-Otwock, Poland
[161] Institute of Mathematics, Polish Academy of Sciences, 00656 Warsaw, Poland
[162] Cornell University, Ithaca, NY 14850, USA
[163] Hillsdale College, Hillsdale, MI 49242, USA
[164] Hanyang University, Seoul 04763, Republic of Korea
[165] Korea Astronomy and Space Science Institute, Daejeon 34055, Republic of Korea
[166] Institute for High-Energy Physics, University of Amsterdam, Science Park 904, 1098 XH Amsterdam, The Netherlands
[167] NASA Marshall Space Flight Center, Huntsville, AL 35811, USA
[168] Dipartimento di Matematica e Fisica, Università degli Studi Roma Tre, I-00146 Roma, Italy
[169] INFN, Sezione di Roma Tre, I-00146 Roma, Italy
[170] ESPCI, CNRS, F-75005 Paris, France
[171] OzGrav, Swinburne University of Technology, Hawthorn VIC 3122, Australia
[172] Southern University and A&M College, Baton Rouge, LA 70813, USA
[173] Centre Scientifique de Monaco, 8 quai Antoine Ier, MC-98000, Monaco
[174] Indian Institute of Technology Madras, Chennai 600036, India
[175] Institut des Hautes Etudes Scientifiques, F-91440 Bures-sur-Yvette, France
[176] IISER-Kolkata, Mohanpur, West Bengal 741252, India
[177] Institut für Kernphysik, Theoriezentrum, D-64289 Darmstadt, Germany

[178] Whitman College, 345 Boyer Avenue, Walla Walla, WA 99362, USA
[179] Université de Lyon, F-69361 Lyon, France
[180] Hobart and William Smith Colleges, Geneva, NY 14456, USA
[181] Department of Physics, Lancaster University, Lancaster LA1 4 YB, UK
[182] Dipartimento di Fisica, Università degli Studi di Torino, I-10125 Torino, Italy
[183] University of Washington, Seattle, WA 98195, USA
[184] Indian Institute of Technology, Gandhinagar Ahmedabad Gujarat 382424, India
[185] INAF, Osservatorio Astronomico di Brera sede di Merate, I-23807 Merate, Lecco, Italy
[186] Centro de Astrofísica e Gravitação (CENTRA), Departamento de Física, Instituto Superior Técnico, Universidade de Lisboa, 1049-001 Lisboa, Portugal
[187] Marquette University, 11420 W. Clybourn St., Milwaukee, WI 53233, USA
[188] Université de Montréal/Polytechnique, Montreal, Quebec H3T 1J4, Canada
[189] Indian Institute of Technology Hyderabad, Sangareddy, Khandi, Telangana 502285, India
[190] INAF, Osservatorio di Astrofisica e Scienza dello Spazio, I-40129 Bologna, Italy
[191] International Institute of Physics, Universidade Federal do Rio Grande do Norte, Natal RN 59078-970, Brazil
[192] Villanova University, 800 Lancaster Ave, Villanova, PA 19085, USA
[193] Andrews University, Berrien Springs, MI 49104, USA
[194] Max Planck Institute for Gravitationalphysik (Albert Einstein Institute), D-14476 Potsdam-Golm, Germany
[195] Università di Siena, I-53100 Siena, Italy
[196] Trinity University, San Antonio, TX 78212, USA
[197] Van Swinderen Institute for Particle Physics and Gravity, University of Groningen, Nijenborgh 4, 9747 AG Groningen, The Netherlands

## Abstract

On 2019 April 25, the LIGO Livingston detector observed a compact binary coalescence with signal-to-noise ratio 12.9. The Virgo detector was also taking data that did not contribute to detection due to a low signal-to-noise ratio, but were used for subsequent parameter estimation. The 90% credible intervals for the component masses range from 1.12 to 2.52 $M_\odot$ (1.46–1.87 $M_\odot$ if we restrict the dimensionless component spin magnitudes to be smaller than 0.05). These mass parameters are consistent with the individual binary components being neutron stars. However, both the source-frame chirp mass $1.44^{+0.02}_{-0.02}\,M_\odot$ and the total mass $3.4^{+0.3}_{-0.1} M_\odot$ of this system are significantly larger than those of any other known binary neutron star (BNS) system. The possibility that one or both binary components of the system are black holes cannot be ruled out from gravitational-wave data. We discuss possible origins of the system based on its inconsistency with the known Galactic BNS population. Under the assumption that the signal was produced by a BNS coalescence, the local rate of neutron star mergers is updated to 250–2810 $\mathrm{Gpc^{-3}\,yr^{-1}}$.

## 1. Introduction

The first observation of gravitational waves from the inspiral of a binary neutron star (BNS)[200] system on 2017 August 17 (Abbott et al. 2017b) was a major landmark in multi-messenger astronomy and astrophysics. The gravitational-wave merger was accompanied by a gamma-ray burst (Abbott et al. 2017c; Goldstein et al. 2017; Savchenko et al. 2017); the subsequent world-wide follow-up of the signal by electromagnetic telescopes and satellite observatories identified the host galaxy and observed the kilonova and afterglow emission of the event over a period of hours to months (see, for example, Abbott et al. 2017d and references therein; Villar et al. 2017; Hajela et al. 2019; Troja et al. 2019).

In this Letter, we present the second observation of a gravitational-wave signal consistent with the inspiral of a BNS system, GW190425. The source properties of this signal imply a total mass and chirp mass larger than any known BNS. There are interesting implications for the formation of this system.

We observed the GW190425 signal on 2019 April 25, 08:18:05 UTC, with it being initially assigned the candidate name S190425z (LIGO Scientific Collaboration & Virgo Collaboration 2019a), during the third observing run (O3) of the LIGO–Virgo network, which started on 2019 April 1. The network consists of two Advanced LIGO interferometers (Aasi et al. 2015) in Hanford, Washington, USA (LHO) and Livingston, Louisiana, USA (LLO) and the Advanced Virgo interferometer in Cascina, Italy (Acernese et al. 2015). At the time of GW190425, LHO was temporarily offline with only LLO and Virgo taking data. GW190425 was detected as a single-detector event in LLO in low latency by the GstLAL-based inspiral search pipeline (Cannon et al. 2012; Privitera et al. 2014; Messick et al. 2017; Hanna et al. 2019; Sachdev et al. 2019). Analyses with three other pipelines also detected a consistent signal. The signal-to-noise ratio (S/N) in Virgo was below the detection threshold. To date, no confirmed electromagnetic or neutrino event has been identified in association with this gravitational-wave event.

## 2. Detectors

Between the second observing run (O2) and O3, several improvements were made to increase the detectors' sensitivity (Aasi et al. 2013; Acernese et al. 2015). For the LIGO detectors, the changes consisted of: the injection of squeezed vacuum at the level of 2–3 dB (Tse et al. 2019); the replacement of the signal recycling mirror with a larger optic with lower transmission; an increase in the input power to about 40 W through the installation of a 70 W amplifier and tuned mass dampers for the high-frequency parametric instabilities of the test masses (Evans et al. 2015; Biscans et al. 2019); the replacement of the end mirrors for lower optical losses; and light baffle installation to mitigate noise from scattered light. The sensitivity, quantified by the angle-averaged BNS inspiral range (see, e.g., the *sense-monitor* range discussion in Allen et al. 2012), was 102–111 Mpc for LHO and 125–140 Mpc for LLO during the first phase of O3. See Abbott et al. (2019a) for sensitivity curves and ranges during O1 and O2, for comparison.

[198] Deceased, 2018 July.
[199] Please direct all correspondence to LSC Spokesperson at lsc-spokesperson@ligo.org, or Virgo Spokesperson at virgo-spokesperson@ego-gw.it.
[200] The term BNS is used here for a system containing two neutron stars, synonymous with the term "double neutron star system" also used in the literature.

For Virgo, the improvements consisted of: the injection of squeezed vacuum at the level of 2–3 dB (Acernese et al. 2019); the replacement of the steel test-mass suspension wires with fused silica fibers; the installation of a 100 W laser amplifier and increase of the interferometer input power from 10 to 18 W; the installation of additional baffles in several critical locations in the interferometer to mitigate scattered light; and the refinement of global alignment control at higher bandwidth than in O2. The Virgo BNS inspiral range was about 43–50 Mpc over the first three months of O3.

At the time of GW190425 only LLO and Virgo were operational; LHO was offline for $\sim$2 hr around the event time. Prior to the signal, LLO had been in a stable operational state for approximately 30 hr, with a BNS inspiral range of $\sim$135 Mpc. Virgo had been in a stable state for approximately 14 hr, with a BNS inspiral range of $\sim$48 Mpc.

The LIGO and Virgo detectors are calibrated by photon pressure from modulated auxiliary lasers inducing test-mass motion (Karki et al. 2016; Acernese et al. 2018; Viets et al. 2018). The maximum $1\sigma$ calibration uncertainties for strain data used in the analysis of GW190425 were 6% in amplitude and 3.5° in phase for LIGO data, and 5% in amplitude and 7° in phase for Virgo data, over the frequency range 19.4–2048 Hz.

We used detection procedures similar to those used to vet previous gravitational-wave events (Abbott et al. 2016a) and found no evidence that environmental or instrumental disturbances (Effler et al. 2015) could account for GW190425. Approximately 60 s prior to the coalescence time of GW190425 there was a short noise transient in LLO. Short noise transients of instrumental origin are common in the LIGO and Virgo detectors. We have verified (see Section 4), that this noise transient does not affect the inference of the signal parameters including the S/N by comparing the signal parameters estimated over the original data to those parameters deduced with a time-frequency wavelet model (Cornish & Littenberg 2015; Pankow et al. 2018; Abbott et al. 2019b) of the noise transient subtracted from the data.

During the first two observing runs, gravitational-wave alerts were sent to partner observatories in order to facilitate multi-messenger astronomy. Starting in O3, these alerts have been made public in low latency and distributed through NASA's Gamma-ray Coordinates Network (GCN).[201]

## 3. Detection of GW190425

We identified GW190425 as a single-detector event in the LLO data using a low-latency matched-filtering search for coalescing binaries, the GstLAL-based inspiral search pipeline (Cannon et al. 2012; Privitera et al. 2014; Messick et al. 2017; Hanna et al. 2019; Sachdev et al. 2019). It was designated the candidate name S190425z in the GraceDB event database.[202] The event had an S/N of 12.9 and an autocorrelation-$\xi^2$ of 0.82 in LLO, with the autocorrelation providing a similar consistency test to a $\chi^2$ value (Messick et al. 2017). Although Virgo was operating at the time of the event, the S/N it observed was only 2.5, which is below the threshold of 4.0 at which searches consider triggers for significance estimation. The difference in S/N between LLO and Virgo is consistent with the difference in the sensitivities of the two detectors. Triggers with consistent S/Ns, signal-consistency-test values, and mass parameters were produced by other low-latency matched-filtering searches, PyCBC Live (Usman et al. 2016; Nitz et al. 2018, 2019), MBTAOnline (Adams et al. 2016), and SPIIR (Hooper et al. 2012; Luan et al. 2012; Chu 2017; Guo et al. 2018) (see Appendix B). The searches used post-Newtonian (PN) waveform models (Blanchet et al. 1995, 2005; Damour et al. 2001; Arun et al. 2009; Buonanno et al. 2009; Blanchet 2014; Mishra et al. 2016) for performing matched-filtering (Sathyaprakash & Dhurandhar 1991; Owen & Sathyaprakash 1999; Harry et al. 2009).

GstLAL ranks all candidates that pass the S/N threshold using the log-likelihood ratio (Cannon et al. 2015) as a detection statistic (Messick et al. 2017; Hanna et al. 2019; Sachdev et al. 2019). The log-likelihood ratio is calculated based on the signal and noise distributions of trigger parameters: S/N, $\xi^2$, the sensitivities of the detectors at the time of the event, and the time and phase delays between the participating interferometers (for coincident triggers). A false alarm rate (FAR) is then assigned to each candidate based on the probability density of the log-likelihood ratio under the noise hypothesis. The background is informed using non-coincident triggers that occur during times when multiple detectors are operating. Log-likelihood ratios assigned to single-detector candidates have larger uncertainties than those of two- or three-detector events, because the background distributions are computed from the triggers of a single detector and cannot be combined with the background from other detectors. This primarily affects the marginally significant triggers that occur at the tail of background distributions, which is poorly resolved. The triggers that are in the bulk and consistent with noise, and the triggers that are well separated from the noise distributions and consistent with signal, can still be identified. An empirically determined parameter, called the penalty, is subtracted from the log-likelihood ratios of all the single-detector candidates, down-weighting their significance, to ensure that only those events with unambigious separation from the background sample are marked as significant. We require this penalty to be such that the single-detector triggers that are well separated from the background samples are still significant even after being penalized, but it should downrank the triggers present at the tail of the background distributions enough that they are recovered as marginally significant candidates at best. The penalty was determined to be 14 for this run based on the results from simulated signals that were injected in the data during non-coincident times.

Following the application of the penalty, GW190425 was identified as a confident detection. The low-latency FAR estimate of the event was one in 69,000 yr. This FAR was estimated using the data collected in O3 up until the time of the event, amounting to 23.5 days. Even though the FAR estimation of single-detector candidates is challenging (Callister et al. 2017), the matched-filter pipelines are capable of identifying loud single-detector events. GW170817 (Abbott et al. 2017b) was initially identified by GstLAL as a single-detector event. To further establish the significance of GW190425, it was compared against the 169.5 days of background from O1 and O2 and 50 days of background from O3 in the BNS part of the parameter space, and found to be louder than any background event. The BNS region is defined as the parameter space with component masses between 1 and 3 $M_\odot$. The results of this background analysis from the GstLAL search are shown in Figure 1, which shows the combined S/N–$\xi^2$ noise probability density function for LHO, LLO, and Virgo. The S/N–$\xi^2$ distributions from O1 and O2 are taken from the analysis performed for GWTC-1 (Abbott et al. 2019c), while the S/N–$\xi^2$ distributions from O3 come from the low-latency search. The S/N–$\xi^2$ background distributions are a subset of the parameters that factor in the calculation of the log-likelihood ratio, which is the detection statistic used by the GstLAL search. These background distributions allow us to include the S/N–$\xi^2$ information from all the triggers, and not just the trigger in question while assigning the detection statistic. Events with low S/Ns and accidentally small residuals would be disfavored by the signal model, which also factors in the log-likelihood ratio.

[201] See the user guide for low-latency alerts at https://emfollow.docs.ligo.org/userguide/.

[202] https://gracedb.ligo.org/

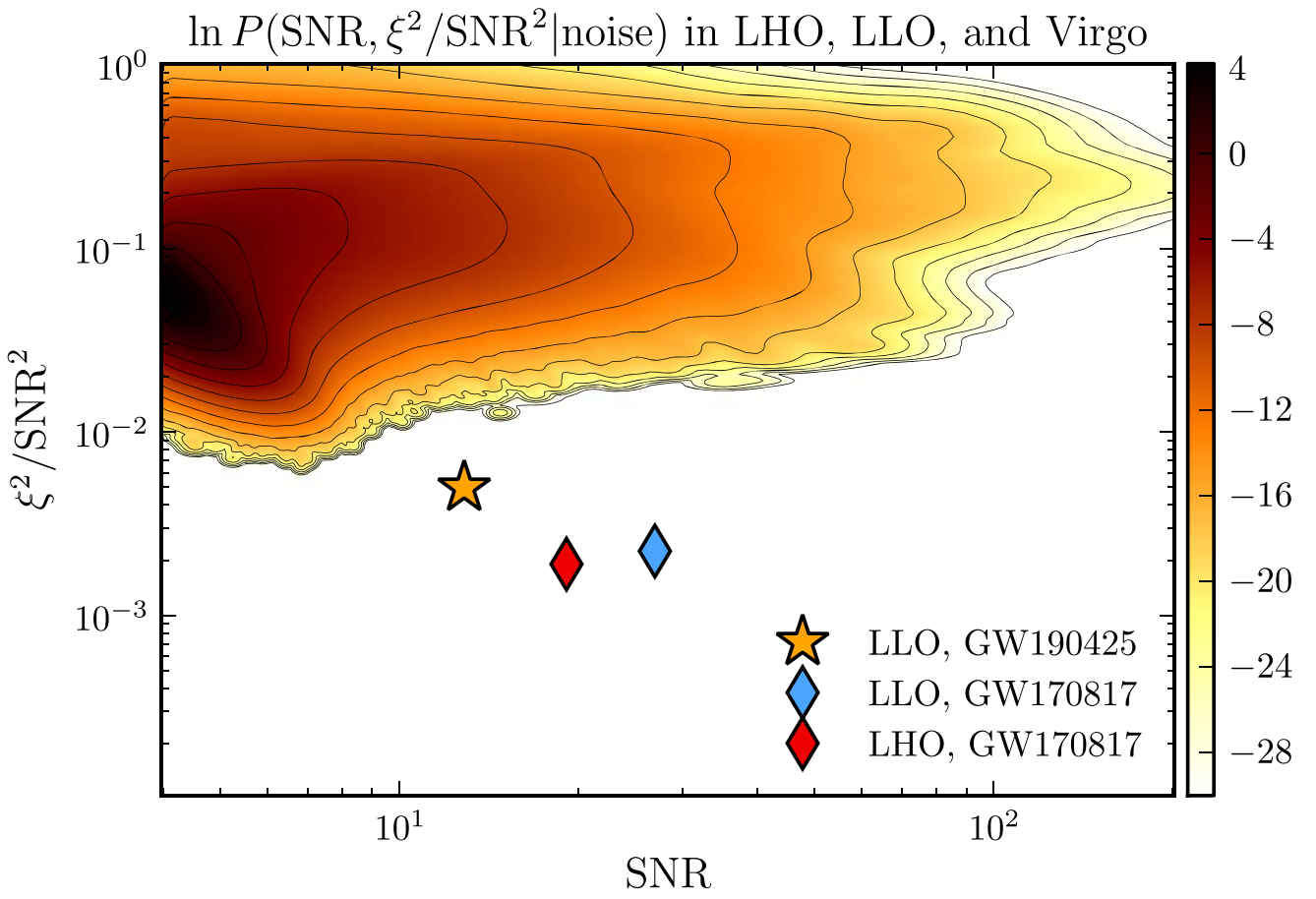

**Figure 1.** Combined S/N–$\xi^2$ noise probability density function for LHO, LLO, and Virgo in the BNS region, computed by adding the normalized 2D histograms of background triggers in the S/N–$\xi^2$/S/N$^2$ plane from the three detectors. The gold star indicates GW190425. There is no background present at the position of GW190425; it stands out above all of the background recorded in the Advanced LIGO and Virgo detectors in the first three observing runs. The background contains 169.5 days of data from O1 and O2 and the first 50 days of O3, at times when any of the detectors were operating. For comparison the LLO and LHO triggers for GW170817 are also shown in the plot as blue and red diamonds, respectively.

As seen in Figure 1, there is no background recorded at the GW190425 parameters in all the data searched over until now. Thus, despite the caveats associated with finding signals in a single detector, GW190425 is a highly significant event that stands out above all background. In Appendix B we also show the results from the PyCBC.

We sent out an alert ∼43 min after the trigger (LIGO Scientific Collaboration & Virgo Collaboration 2019a), which included a sky map computed using a rapid Bayesian algorithm (Singer & Price 2016). We assigned GW190425 a >99% probability of belonging to the BNS source category. The initial sky map had a 90% credible region of 10,200 deg$^2$. Although data from both LLO and Virgo were used to constrain the sky location, it extended over a large area due to the fact that the signal was only observed with high confidence in a single observatory. Gravitational-wave localization relies predominantly on measuring the time delay between observatories. However, in this case it is primarily the observed stain amplitude that localizes the signal, with the more likely parts of the sky being dominated by positions where the the antenna response of LLO is favorable.

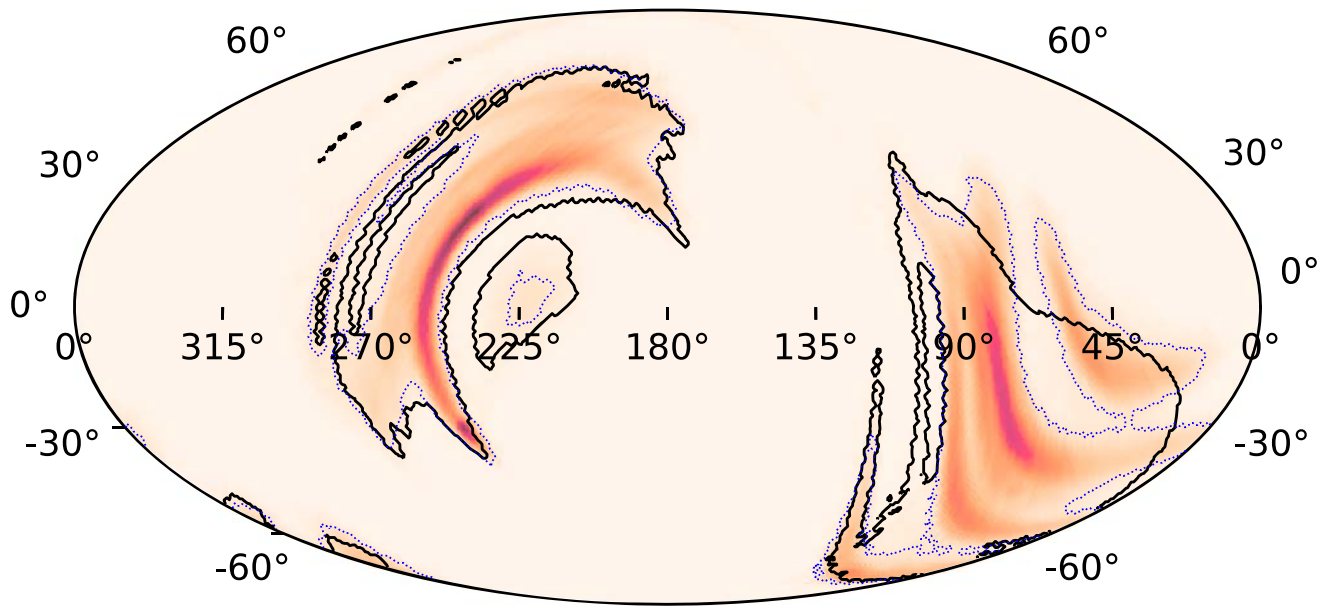

**Figure 2.** Sky map for GW190425. The shaded patch is the sky map obtained from the Bayesian parameter estimation code LALInference (Veitch et al. 2015) (see Section 4) with the 90% confidence region bounded by the thin dotted contour. The thick solid contour shows the 90% confidence region from the low-latency sky localization algorithm BAYESTAR (Singer & Price 2016).

We generated an improved sky map using a Bayesian analysis that sampled over all binary system parameters (see Section 4), producing a 90% credible sky area of 8284 deg$^2$ and a distance constrained to $159^{+69}_{-71}$ Mpc. This sky map, and the initial low-latency map, are shown in Figure 2. As a comparison, GW170817 was localized to within 28 deg$^2$ at a 90% credible level. The broad probability region in the sky map for this event presented a significant challenge for follow-up searches for electromagnetic counterparts. At the time of writing, no clear detection of a counterpart has been reported in coincidence with GW190425 (e.g., Coughlin et al. 2019; Hosseinzadeh et al. 2019; Lundquist et al. 2019, but also see Pozanenko et al. 2019), although a wide range of searches for coincident electromagnetic or neutrino signals have been performed and reported in the GCN Circular archive.[203]

## 4. Source Properties

We have inferred the parameters of the GW190425 source using a coherent analysis of the data from LLO and Virgo (in the frequency range 19.4–2048 Hz) following the methodology described in Appendix B of Abbott et al. (2019c).[204] The low-frequency cutoff of 19.4 Hz was chosen such that the signal was in-band for the 128 s of data chosen for analysis. In this frequency range there were ∼3900 phase cycles before merger.

We cleaned the data from LLO to remove lines from calibration and from known environmental artifacts (Davis et al. 2019; Driggers et al. 2019). For Virgo, we used the low-latency data. The LLO data were subsequently pre-processed (Cornish & Littenberg 2015; Pankow et al. 2018) to remove the noise transient discussed in Section 2. Details of the transient model and the data analyzed can be found in Abbott et al. (2019b). The results have been verified to be robust to this glitch removal by comparing the analysis of the pre-processed data with that using the non-pre-processed data and by comparing results with a low-frequency cutoff of 30 Hz. We estimated the noise spectra of the data from both detectors using the methods described in Littenberg & Cornish (2015) and Chatziioannou et al. (2019).

[203] All GCN Circulars related to this event are archived at https://gcn.gsfc.nasa.gov/other/S190425z.gcn3.

[204] From here on, we will use GW190425 to refer to the gravitational-wave signal and as shorthand for the system that produced the signal.

We estimated the posterior probability distribution for the source model parameter space using the Bayesian stochastic sampling software in LALInference (Veitch et al. 2015); the analysis marginalized over the uncertainty in detector calibration (Cahillane et al. 2017). The data used in this analysis are open-access and available from the Gravitational Wave Open Science Centre (LIGO Scientific Collaboration & Virgo Collaboration 2019b).

The primary analysis presented here was produced using the PhenomPv2NRT signal model (Dietrich et al. 2019), a phenomenological waveform model for spin-precessing (Hannam et al. 2014; Khan et al. 2016) compact binary systems, which also includes tidal interactions (Dietrich et al. 2017). At the S/N of GW190425, it is not expected that systematic errors coming from our choice of waveform approximant would be significant. Indeed, comparisons between PhenomPv2NRT and effective-one-body (EOB) tidal models (Hinderer et al. 2016; Nagar et al. 2018) in the case of GW170817 suggested that even at the relatively high S/N of 33, model systematics were subdominant to statistical errors (Abbott et al. 2019c). To verify this expectation, we also obtained results with three further models: SEOBNRv4Tsurrogate (Hinderer et al. 2016; Steinhoff et al. 2016; Bohé et al. 2017; Lackey et al. 2019), IMRPhenomDNRT (Husa et al. 2016; Khan et al. 2016; Dietrich et al. 2017, 2019), and TaylorF2 (Sathyaprakash & Dhurandhar 1991; Poisson 1998; Mikóczi et al. 2005; Arun et al. 2009; Bohé et al. 2013, 2015; Mishra et al. 2016) and conclude that our findings are robust with respect to waveform systematics. We present details of this investigation in Appendix D. For the PhenomPv2NRT and PhenomDNRT waveforms, we applied the reduced-order quadrature method for evaluating the likelihood (Smith et al. 2016; Baylor et al. 2019; Smith 2019) to reduce the overall computational cost.

We chose a uniform prior between 1.00 $M_\odot$ and 5.31 $M_\odot$ for the redshifted (detector-frame) component masses and used the conventional definition that $m_1 \geqslant m_2$. As in Abbott et al. (2019d), we present separate results from using different low-spin and high-spin priors, with dimensionless spin magnitudes ($\chi = |\boldsymbol{\chi}|$) for both components uniformly distributed within $\chi < 0.05$ and $\chi < 0.89$, respectively, and assuming that the spin directions are isotropically distributed. The low-spin prior was chosen so as to include the fastest pulsars among known Galactic BNS systems that will merge within a Hubble time (Zhu et al. 2018) although, as we show below, for this event the chirp mass is not consistent with the known Galactic BNS systems. We gave the component tidal deformability parameters uniform priors in the ranges $\Lambda_1 \in [0, 5\,000]$ and $\Lambda_2 \in [0, 10\,000]$; the distinct prior ranges were selected to ensure that the priors did not affect regions with significant posterior support. These prior ranges are consistent with the constraints imposed by causality (Van Oeveren & Friedman 2017).

All results below are given assuming the high-spin prior unless otherwise stated. The secondary mass $m_2$ has posterior support near to the arbitrary bounds enforced by the reduced-order quadrature method for PhenomPv2NRT. However, results from the TaylorF2 waveform, with a lower prior bound on $m_2$ of 0.7, confirm that these restrictions do not affect the overall results.

**Table 1**
Source Properties for GW190425

| | Low-spin Prior ($\chi < 0.05$) | High-spin Prior ($\chi < 0.89$) |
|---|---|---|
| Primary mass $m_1$ | 1.60–1.87 $M_\odot$ | 1.61–2.52 $M_\odot$ |
| Secondary mass $m_2$ | 1.46–1.69 $M_\odot$ | 1.12–1.68 $M_\odot$ |
| Chirp mass $\mathcal{M}$ | $1.44^{+0.02}_{-0.02}$ $M_\odot$ | $1.44^{+0.02}_{-0.02}$ $M_\odot$ |
| Detector-frame chirp mass | $1.4868^{+0.0003}_{-0.0003}$ $M_\odot$ | $1.4873^{+0.0008}_{-0.0006}$ $M_\odot$ |
| Mass ratio $m_2/m_1$ | 0.8 – 1.0 | 0.4 – 1.0 |
| Total mass $m_{\rm tot}$ | $3.3^{+0.1}_{-0.1}$ $M_\odot$ | $3.4^{+0.3}_{-0.1}M_\odot$ |
| Effective inspiral spin parameter $\chi_{\rm eff}$ | $0.012^{+0.01}_{-0.01}$ | $0.058^{+0.11}_{-0.05}$ |
| Luminosity distance $D_{\rm L}$ | $159^{+69}_{-72}$ Mpc | $159^{+69}_{-71}$ Mpc |
| Combined dimensionless tidal deformability $\tilde{\Lambda}$ | $\leqslant$600 | $\leqslant$1100 |

**Note.** We give ranges encompassing the 90% credible intervals for the PhenomPv2NRT model; in Appendix D we demonstrate these results are robust to systematic uncertainty in the waveform. Mass values are quoted in the frame of the source, accounting for uncertainty in the source redshift. For the primary mass we give the 0%–90% interval, while for the secondary mass and mass ratio we give the 10%–100% interval: the uncertainty on the luminosity distance means that there is no well-defined equal-mass bound for GW190425. The quoted 90% upper limits for $\tilde{\Lambda}$ are obtained by reweighting its posterior distribution as detailed in Appendix F.1.

In Table 1 we summarize the inferred values for a selection of the source parameters; unless otherwise stated, all bounds are given by a 90% credible interval, symmetric in probability about the median of the marginalized posterior probability distribution for a given parameter. Frequency-dependent binary parameters are quoted at 20 Hz.

Assuming a standard flat ΛCDM cosmology with Hubble constant $H_0 = 67.9\ \mathrm{km\ s^{-1}\ Mpc^{-1}}$ and matter density parameter $\Omega_m = 0.306$ (Ade et al. 2016), we infer the cosmological redshift to be $z = 0.03^{+0.01}_{-0.02}$. The redshift from peculiar velocity is expected to be negligible (see Carrick et al. 2015). Therefore, we find the source-frame chirp mass to be $\mathcal{M} = 1.44^{+0.02}_{-0.02}\,M_\odot$. From the source-frame chirp mass and inferred mass ratio, we constrain the primary mass to the range $[1.61, 2.52]\,M_\odot$ and the secondary mass to the range $[1.12, 1.68]\,M_\odot$ as shown in Figure 3. We discuss the implications of the chirp mass and the total system mass of $3.4^{+0.3}_{-0.1}M_\odot$ in Section 5.

Spin effects are measurable primarily through the effective spin parameter $\chi_{\rm eff}$ (Racine 2008; Ajith et al. 2011), which is the mass-weighted sum of spins projected along the direction perpendicular to the orbital plane. In Figure 4 we show the joint posterior distribution between $\chi_{\rm eff}$ and mass ratio ($q = m_2/m_1$) along with one-dimensional posterior distributions. The $\chi_{\rm eff}$–$q$ correlation causes a positive skew in the marginalized $\chi_{\rm eff}$ posterior (Cutler & Flanagan 1994). To quantify the support for spins in GW190425, we calculated the Bayesian evidence for the same PhenomPv2NRT model, but with spin effects turned off. We found a Bayes factor of ∼1 between the non-spinning and spinning cases, implying no evidence for or against spins. In order to place Figure 4 in an astrophysical context, we also show the mass ratios and expected effective spins at merger for the two fastest Galactic BNS systems that are expected to merge within a Hubble time. For the double pulsar J0737−3039A/B, precise mass and spin-period measurements are available for both components (Kramer et al. 2006). With a mass ratio of 0.93, it is expected to have $\chi_{\rm eff}$ between 0.008 and 0.012 (90% credibility interval) when marginalized over mass and equation of state (EoS) uncertainties (see Appendix F.3 for details). The fastest-spinning Galactic-field BNS, which contains the 17 ms pulsar J1946+2052 (Stovall et al. 2018), has $\chi_{\rm eff}$ in the range [0.012, 0.018] assuming aligned spin for the pulsar and negligible spin for its companion, similar to the double pulsar.

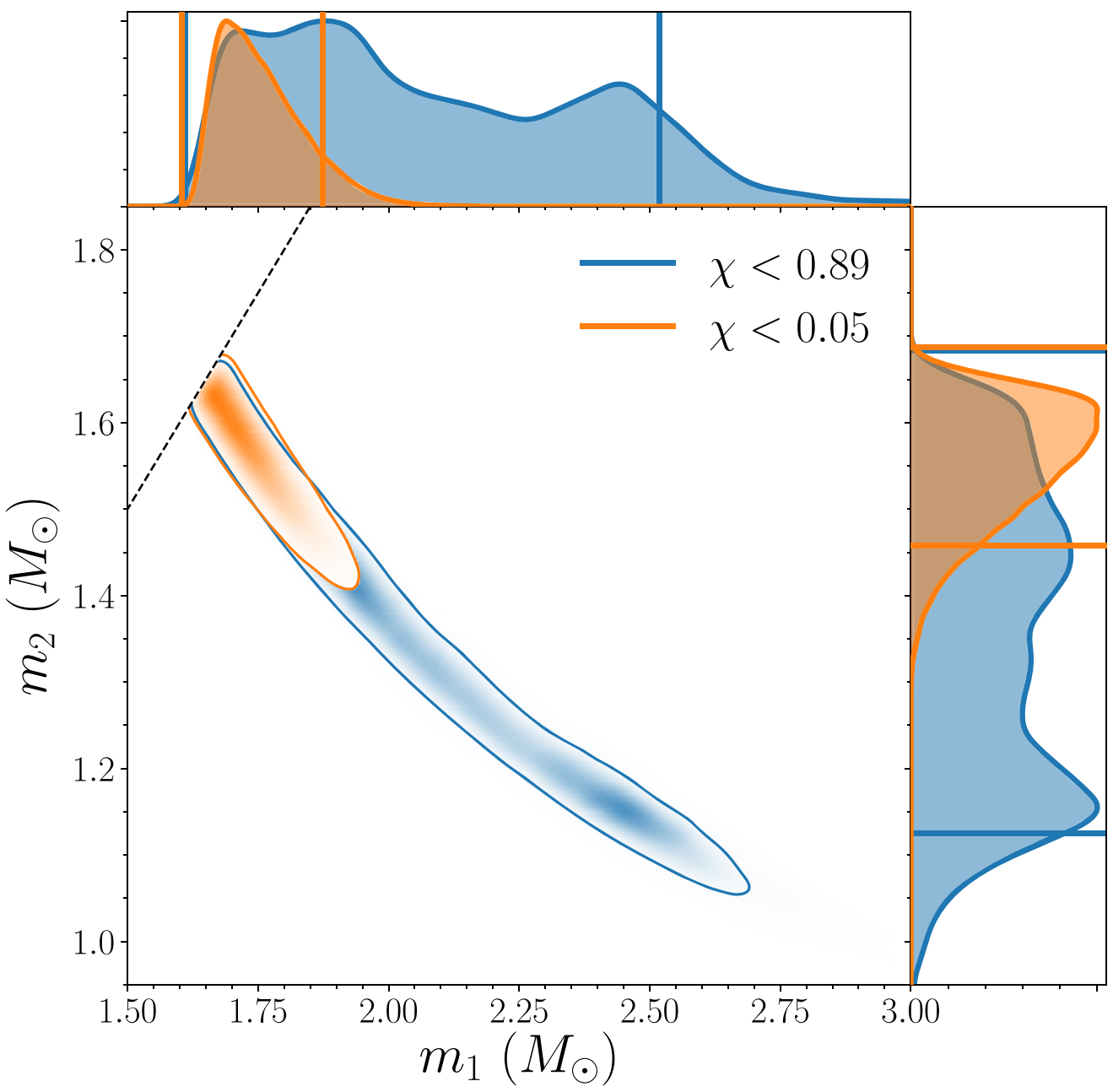

**Figure 3.** Posterior distribution of the component masses $m_1$ and $m_2$ in the source frame for the low-spin ($\chi < 0.05$; orange) and high-spin ($\chi < 0.89$; blue) analyses. Vertical lines in the one-dimensional plots enclose 90% of the probability and correspond to the ranges given in Table 1. The one-dimensional distributions have been normalized to have equal maxima. A dashed line marks the equal-mass bound in the two-dimensional plot.

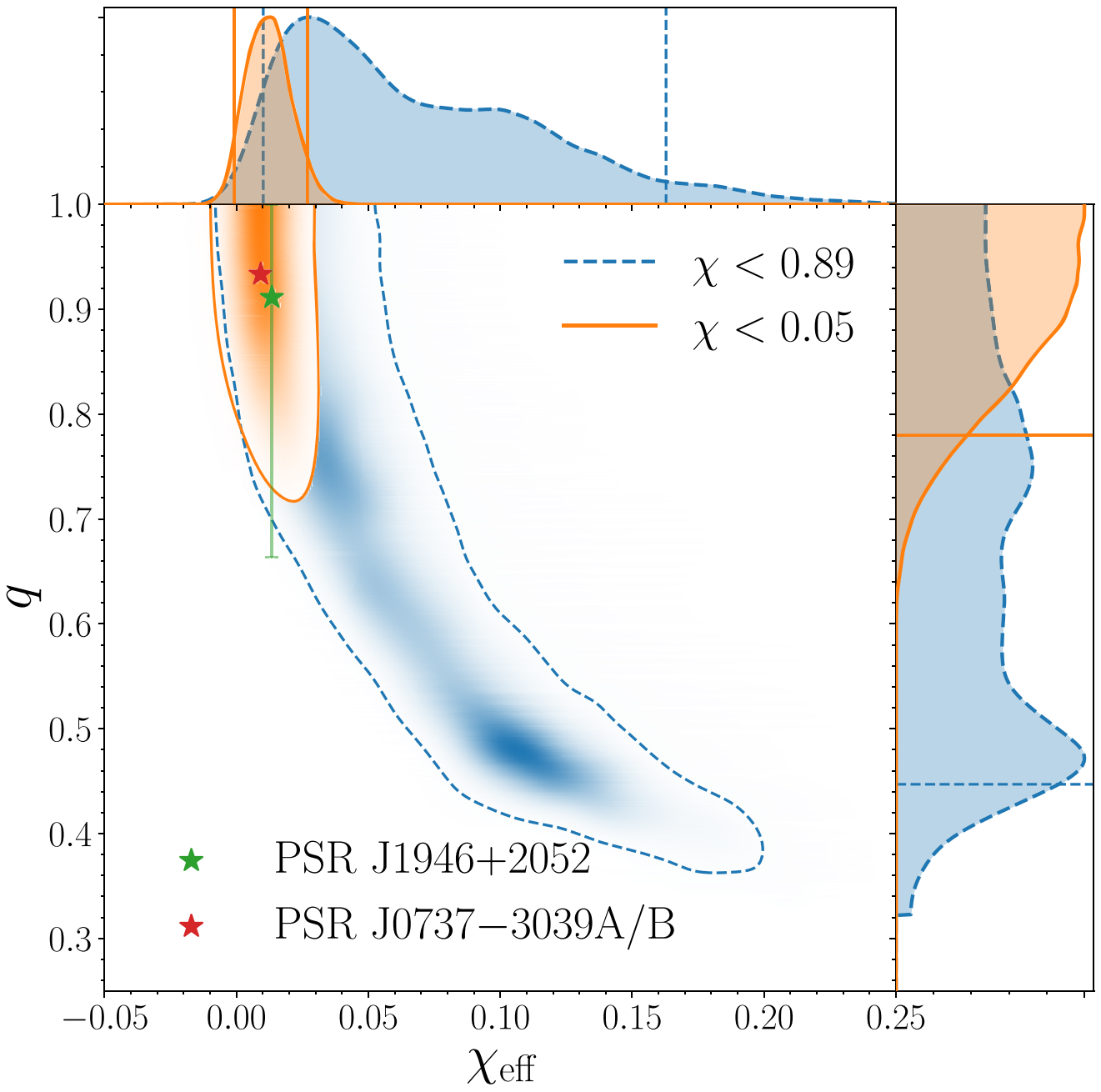

**Figure 4.** Joint posterior distribution of $\chi_{\rm eff}$ and $q$ for the low-spin ($\chi < 0.05$; orange) and high-spin ($\chi < 0.89$; blue) prior. Vertical lines enclose the 90% credible interval for $\chi_{\rm eff}$ and horizontal lines mark the 90% lower limits for $q$. The one-dimensional distributions have been normalized to have equal maxima. For comparison, the effective spins are shown for two Galactic BNS systems, PSR J1946+2052 (green) and PSR J0737−3039A/B (red), if extrapolated to their mergers. For PSR J1946+2052, it is assumed that the primary spin is perpendicular to the orbital plane and that the unmeasured secondary spin is negligible. Uncertainties in the pulsar $q$ and $\chi_{\rm eff}$ values, calculated by marginalizing over mass and equation of state information, are smaller than the markers except for the mass ratio of PSR J1946+2052, which is shown with an error bar.

For the results reported herein we used the LALInference library's nested sampling algorithm and validated results using the LALInference Markov chain Monte Carlo sampling algorithm and the Bilby (Ashton et al. 2019) library with the Dynesty (Speagle 2019) nested sampling algorithm. When comparing the high-spin prior results using the different algorithms, we see $\lesssim$3% differences in the median parameter values and the credible intervals are consistent and reproducible. Meanwhile, the runs using the low-spin priors show no such differences.

We show the posteriors for a wider range of source parameters in Appendix C.

### 4.1. Neutron Star Matter

Because of its large mass, the discovery of GW190425 suggests that gravitational-wave analyses can access densities several times above nuclear saturation (see, e.g., Figure 4 in Douchin & Haensel 2001) and probe possible phase transitions inside the core of a neutron star (NS) (Oertel et al. 2017; Essick et al. 2019; Tews et al. 2019). However, binaries comprised of more massive stars are described, for a fixed EoS, by smaller values of the leading-order tidal contribution to the gravitational-wave phasing $\tilde{\Lambda}$ (Flanagan & Hinderer 2008). These are intrinsically more difficult to measure. For GW190425, this is exacerbated by the fairly low S/N of the event compared to GW170817. Overall, we find that constraints on tides, radius, possible *p–g* instabilities (Venumadhav et al. 2013; Weinberg et al. 2013; Weinberg 2016; Zhou & Zhang 2017), and the EoS from GW190425 are consistent with those obtained from GW170817 (Abbott et al. 2017b, 2019e). However, GW190425 is less constraining of NS properties, limiting the radius to only below 15 km, $\tilde{\Lambda}$ to below 1100 and only ruling out phenomenological *p–g* amplitudes above 1.3 times the 90% upper limit obtained from GW170817 at the same confidence level. The *p–g* constraints were obtained with a different high-spin prior than the rest of the results (see Appendix F.5) but the difference does not significantly change our conclusions. Spin priors can affect the inference of tidal and EoS parameters, and we note that the low-spin results are generally more constrained. Following Agathos et al. (2020), we estimate the probability of the binary promptly collapsing into a black hole (BH) after merger to be 96%, with the low-spin prior, or 97% with the high-spin prior. Repeating the analyses of Chatziioannou et al. (2017) and Abbott et al. (2019d), we find no evidence of a postmerger signal in the 1 s of data surrounding the time of coalescence. We obtain 90% credible upper limits on the strain amplitude spectral density and the energy spectral density of $1.1 \times 10^{-22}\ \mathrm{Hz}^{-1/2}$ and $0.11\ M_\odot\ c^2\ \mathrm{Hz}^{-1}$, respectively, for a frequency of 2.5 kHz. Similar to GW170817, this upper limit is higher than any expected post-merger emission from the binary (Abbott et al. 2019d). More details on all calculations and additional analyses are provided in Appendix F.7.

## 5. Astrophysical Implications

The component masses of GW190425 are consistent with mass measurements of NSs in binary systems (Antoniadis et al. 2016; Alsing et al. 2018) as well as expected NS masses in supernova explosion simulations (Woosley et al. 2002; Burrows et al. 2019; Ebinger et al. 2019a, 2019b). Taking a fiducial range of NS masses between 1.2 and 2.3 $M_{\odot}$, our low-spin posteriors are entirely consistent with both objects being NSs, while there is $\sim$25% of posterior support for component masses outside this range given the high-spin prior. The lower end of this fiducial range corresponds to the lowest precisely measured NS mass, $1.174 \pm 0.004\, M_{\odot}$ for the companion of PSR J0453+1559 in Martinez et al. (2015) (see Tauris & Janka 2019 for an alternative white-dwarf interpretation). It is also difficult to form light NSs with masses below $\sim 1.2\, M_{\odot}$ in current supernova explosion simulations (Burrows et al. 2019; Müller et al. 2019). The upper end is based on the highest precise NS mass measurement of $2.14^{+0.20}_{-0.18}\, M_{\odot}$ (95% credibility interval) for PSR J0740+6620 in Cromartie et al. (2019; see also Abbott et al. 2020 for a discussion of NS upper mass bounds).

Here we discuss the implications for the GW190425 system origin assuming it consists of a pair of NSs. Under this assumption, we have calculated the astrophysical rate of merger when including GW190425. We also briefly discuss the possibility of the system containing BH components.

### 5.1. Possible System Origins

Currently there are 17 known Galactic BNSs with total mass measurements, ranging from 2.50 to 2.89 $M_{\odot}$; 12 of them have masses measured for both components, implying chirp masses from 1.12 to 1.24 $M_{\odot}$ (see Table 1 in Farrow et al. 2019 and references therein for details). In order to quantify how different the source of GW190425 is from the observed Galactic population, we fit the total masses of the 10 binaries that are expected to merge within a Hubble time with a normal distribution. This results in a mean of 2.69 $M_{\odot}$ and a standard deviation of 0.12 $M_{\odot}$. With a total mass of $3.4^{+0.3}_{-0.1} M_{\odot}$, GW190425 lies five standard deviations away from the known Galactic population mean (see Figure 5).[205] A similar ($\gtrsim 5\sigma$) deviation is found if we compare its chirp mass to those of Galactic BNSs. This may indicate that GW190425 formed differently than known Galactic BNSs.

There are two canonical formation channels for BNS systems: the isolated binary evolution channel (Flannery & van den Heuvel 1975; Massevitch et al. 1976; Smarr & Blandford 1976; for reviews see Kalogera et al. 2007; Postnov & Yungelson 2014), and the dynamical formation channel (see Phinney & Sigurdsson 1991; Prince et al. 1991; Grindlay et al. 2006; Lee et al. 2010; Ye et al. 2019, and references therein). The former is the standard formation channel for Galactic-field BNSs (e.g., Tauris et al. 2017), in which the two NSs are formed in a sequence of supernova explosions that occur in an isolated binary.

Assuming a formation through the standard channel, GW190425 might suggest a population of BNSs formed in ultra-tight orbits with sub-hour orbital periods. Such binaries are effectively invisible in current radio pulsar surveys due to severe Doppler smearing (Cameron et al. 2018) and short inspiral times ($\lesssim$10 Myr), but have been predicted to exist in theoretical studies (e.g., Belczynski et al. 2002; Dewi & Pols 2003; Ivanova et al. 2003), and possibly with a comparable formation rate to the currently observed Galactic sample (Vigna-Gómez et al. 2018). The formation of GW190425's source might have involved a phase of stable or unstable mass transfer from a post-helium main-sequence star onto the NS. If the mass ratio between the helium-star donor and the NS were high enough, the mass transfer would be dynamically unstable and lead to a Case BB common-envelope phase that could significantly shrink the binary orbit to sub-hour periods (Ivanova et al. 2003; Tauris et al. 2017). If it is possible for a binary to survive this common envelope phase, the high mass of GW190425 may be indicative of this formation pathway, since a more massive helium-star progenitor of the second-born NS would be required for a common envelope to form. In this process the secondary would likely be ultra-stripped, and so the subsequent supernova kick may be suppressed (Tauris et al. 2015). The small supernova kick, combined with the very tight orbital separation, will increase the probability that the binary remained bound following the supernova that formed the BNS. Additionally, the high mass of GW190425 may point to its NSs being born from low-metallicity stars (e.g., Ebinger et al. 2019b). Giacobbo & Mapelli (2018) showed that BNSs with total masses of 3.2–3.5 $M_{\odot}$ can be formed from isolated binaries provided that the metallicity is relatively low ($\sim$5%–10% solar metallicity). Athough not obviously related to scenarios discussed here, the high-mass X-ray binary Vela X-1 contains an NS with varying mass estimates from 1.5 up to 2.1 $M_{\odot}$ (Barziv et al. 2001; Quaintrell et al. 2003; Falanga et al. 2015; Giménez-García et al. 2016) in a nine day orbit with a $\sim$22 $M_{\odot}$ supergiant star companion. Though it is unlikely that the Vela X-1 system will survive a future common envelope phase (Belczynski et al. 2012), if it does survive the supergiant will eventually undergo core collapse forming an NS or BH, potentially leading to a high-mass BNS similar to GW190425. The existence of a fast-merging channel for the formation of BNSs could be detected by future space-based gravitational-wave detectors (Andrews et al. 2019; Lau et al. 2020).

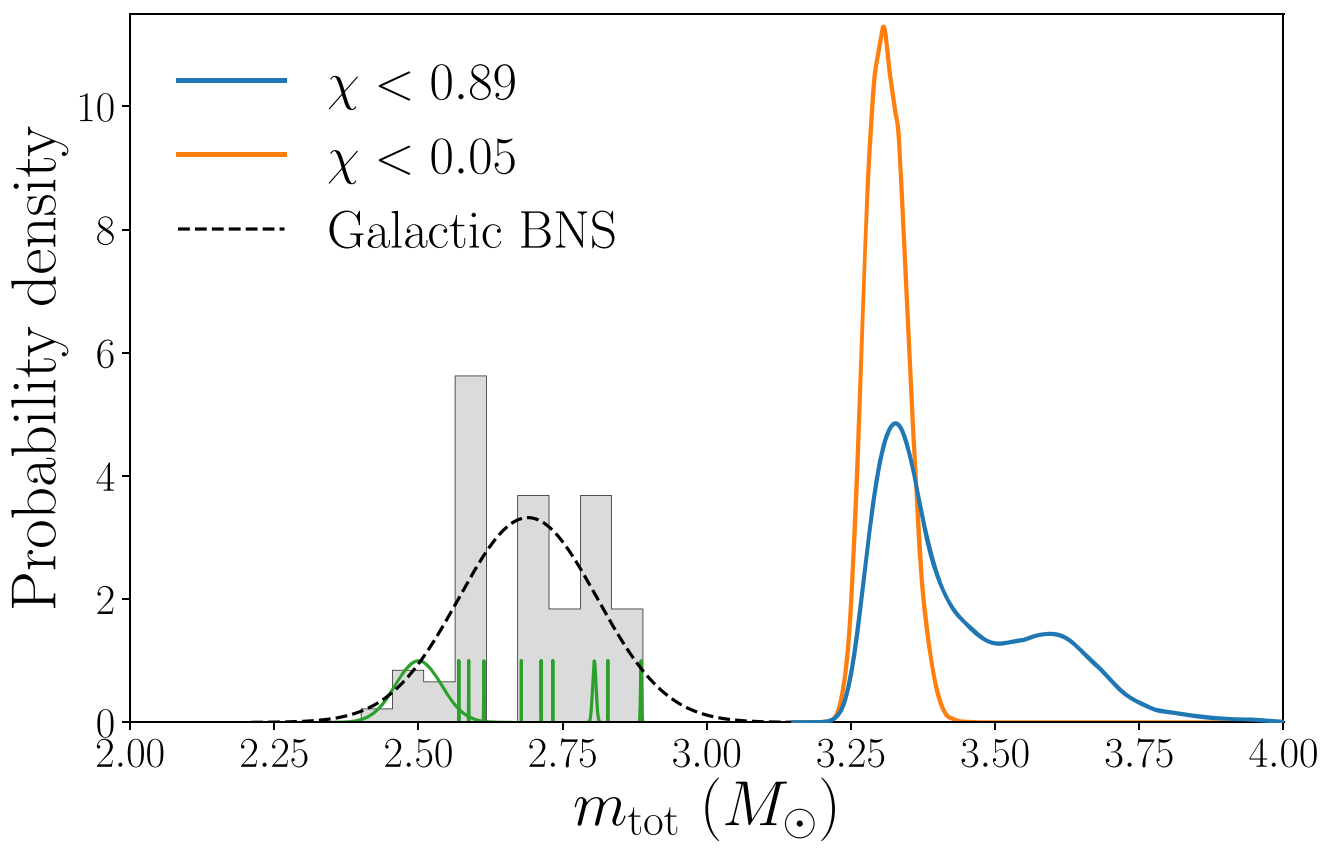

**Figure 5.** Total system masses for GW190425 under different spin priors, and those for the 10 Galactic BNSs from Farrow et al. (2019) that are expected to merge within a Hubble time. The distribution of the total masses of the latter is shown and fit using a normal distribution shown by the dashed black curve. The green curves are for individual Galactic BNS total mass distributions rescaled to the same ordinate axis height of 1.

An alternative way to make the GW190425 system is to have the stellar companion of a massive NS replaced with another NS through a dynamical encounter. Observations of millisecond pulsars in globular clusters have found evidence of massive NSs up to $\sim 2 M_\odot$ (Ransom et al. 2005; Freire et al. 2008). Current modeling of globular clusters suggest that the dynamical formation channel has a negligible contribution to the BNS merger rate in the local universe (Belczynski et al. 2018; Ye et al. 2019), which makes a dynamical origin for GW190425 unlikely. However, a dynamical formation scenario was recently proposed for three BNSs in the Galactic field with similarly short orbital periods and high eccentricities to the globular-cluster BNS PSR B2127+11C (Jacoby et al. 2006); Andrews & Mandel (2019) argued that the clustering of these binaries in the orbital period–eccentricity space challenges the standard binary evolution theory, and further proposed that they were formed in globular clusters, but ejected into the field due to dynamical interactions.

[205] PSR J2222−0137, with a mass of $1.76 \pm 0.06\, M_{\odot}$, is also in a high-mass binary (with $m_{\rm tot} = 3.05 \pm 0.09\, M_{\odot}$, $3\sigma$ higher than the mean of the Galactic BNS population, Cognard et al. 2017); however, the secondary is believed to be a white dwarf rather than an NS.

Another explanation for the large primary mass of GW190425 is that the event is gravitationally lensed, coming from a lower-mass source at higher redshift (Wang et al. 1996; Dai et al. 2017; Hannuksela et al. 2019). However this is highly unlikely considering standard estimates of merger rate evolution and lensing optical depth (Ng et al. 2018; Oguri 2018).

### 5.2. Astrophysical Rate

In Abbott et al. (2019c), the BNS merger rate $\mathcal{R}$ was found to be 110–2520 $\mathrm{Gpc^{-3}\,yr^{-1}}$ assuming a uniform (0.8–2.3 $M_\odot$) component-mass distribution. Here we calculate two rates, alternatively treating GW170817 and GW190425 as two counts from this same uniform-in-component-mass population, or as one count each from GW170817-like and GW190425-like populations. We calculated the sensitive volume of these two mass models semi-analytically, setting a nominal once per century FAR threshold and calibrating to results of the GstLAL search pipeline run on injected signals.

Taking the uniform component-mass distribution from Abbott et al. (2019c), counting both BNS events as two detections during O1, O2, and 50 days of O3, and applying an $\mathcal{R}^{-1/2}$ Jeffreys prior, gives a BNS merger rate of $980^{+1490}_{-730}\ \mathrm{Gpc^{-3}\,yr^{-1}}$. Alternatively, using the method of Kim et al. (2003) as previously used in Abbott et al. (2016b), we have also calculated both GW170817-like and GW190425-like merger rates according to our sensitivities during O1, O2, and 50 days of O3 to BNS populations with the inferred mass and spin distributions for GW170817 and GW190425, respectively. These give $\mathcal{R}_{170817} = 760^{+1740}_{-650}\ \mathrm{Gpc^{-3}\,yr^{-1}}$ and $\mathcal{R}_{190425} = 460^{+1050}_{-390}\ \mathrm{Gpc^{-3}\,yr^{-1}}$. Combining these according to Kim et al. (2003) forms a total BNS rate $\mathcal{R} = \mathcal{R}_{170817} + \mathcal{R}_{190425}$ and, after applying the same Jeffreys prior, gives a BNS merger rate of $1090^{+1720}_{-800}\ \mathrm{Gpc^{-3}\,yr^{-1}}$.

Both estimates are broadly consistent with previous BNS merger rates. The inferred lower limits are higher than the previous estimate, and potentially in tension with the lower BNS merger rates predicted by multiple studies (Chruslinska et al. 2018; Kruckow et al. 2018; Mapelli & Giacobbo 2018; Eldridge et al. 2019), but they are consistent with the most recent population synthesis models (Bray & Eldridge 2018; Giacobbo & Mapelli 2019). These are also consistent with the rates estimated from observations of the Galactic BNS population when taking into account the range of systematic uncertainties (Pol et al. 2019).

### 5.3. Black Holes

A BNS merger is the most straightforward explanation for GW190425. However, the possibility that one or both binary components of GW190425 are BHs cannot be ruled out with a gravitational-wave analysis because we lack the requisite sensitivity to detect matter effects.

A BH interpretation of GW190425 would require BHs falling in the apparent mass gap between NSs and BHs (Bailyn et al. 1998; Özel et al. 2010; Farr et al. 2011), the existence of which is still under debate. Some theoretical models of supernova explosions predicted a smooth transition from NS to BH masses (Woosley & Weaver 1995; Fryer & Kalogera 2001; Ertl et al. 2020), while others suggested a lower limit of BH masses at $\sim 4\ M_\odot$ (Kochanek 2014; Pejcha & Thompson 2015). In addition to supernovae remnants, it is also possible to fill the mass gap with BNS merger remnants (e.g., Gupta et al. 2019). Kreidberg et al. (2012) argued that the mass gap can be explained by possible systematic errors in the mass measurements of BHs in X-ray binaries. Recently, Wyrzykowski & Mandel (2019) found that the mass gap is disfavored by microlensing measurements of *Gaia* Data Release 2 if small BH natal kicks (<20–80 $\mathrm{km\,s^{-1}}$) are assumed.

As an alternative to stellar-origin BHs, a more exotic possibility is that GW190425 was a merger of primordial black holes (PBHs). It has been speculated that PBHs may make up the binaries detected by gravitational-wave detectors (Bird et al. 2016; Sasaki et al. 2016; Clesse & García-Bellido 2017). Byrnes et al. (2018) have shown that if PBHs were produced in the mass range relevant to gravitational-wave detectors, their mass function should consist of a peak around one solar mass. In the scenario that GW190425 was produced by the merger of PBHs, the implied merger rate would then be consistent with the upper limits for subsolar-mass BH mergers (Abbott et al. 2019f) and the possibility that one or more of the previously detected BH mergers are of primordial origin.

## 6. Conclusions

GW190425 represents a highly significant gravitational-wave signal most likely originating from the merger of two NSs, which would make it the second such signal to be observed with gravitational waves. The low-latency FAR for the signal, as estimated by the GstLAL pipeline, was one in 69,000 yr. The signal only passed the detection threshold in a single detector, but it has a detection statistic that was a distinct outlier from the single detector triggers seen in the previous O1 and O2 observing runs.

If the source of GW190425 is a BNS system, it is significantly different from the known population of Galactic double NS systems, with a total mass ($3.4^{+0.3}_{-0.1} M_\odot$) and chirp mass ($1.44^{+0.02}_{-0.02}\ M_\odot$) larger than any of the Galactic systems (Farrow et al. 2019; Zhang et al. 2019). This may have implications for the system's origin, suggesting isolated formation in ultra-tight orbits with sub-hour orbital periods, or formation through the dynamical channel. Since we cannot see evidence of tides, it is possible that one or both objects could be BHs. However, this would require a previously unaccounted-for formation channel for binary BHs in this mass range. In the BNS scenario, the detection of GW190425 provides an update on the rate of BNS mergers of 250–2810 $\mathrm{Gpc^{-3}\,yr^{-1}}$, taking the union between two scenarios for component mass distributions. In either case, the source of GW190425 represents a previously undetected type of astrophysical system. Future gravitational-wave observations of BNS mergers and electromagnetic follow-ups should greatly improve our understanding of BNS formation.

Stretches of data containing this signal, and samples from the posterior probability distributions of the source parameters, are available from the Gravitational Wave Open Science Center (LIGO Scientific Collaboration & Virgo Collaboration 2019b). The software packages used in our analysis are open source.

The authors gratefully acknowledge the support of the United States National Science Foundation (NSF) for the construction and operation of the LIGO Laboratory and Advanced LIGO as well as the Science and Technology Facilities Council (STFC) of the United Kingdom, the Max-Planck-Society (MPS), and the State of Niedersachsen/Germany for support of the construction of Advanced LIGO and construction and operation of the GEO600 detector. Additional support for Advanced LIGO was provided by the Australian Research Council. The authors gratefully acknowledge the Italian Istituto Nazionale di Fisica Nucleare (INFN), the French Centre National de la Recherche Scientifique (CNRS) and the Foundation for Fundamental Research on Matter supported by the Netherlands Organisation for Scientific Research, for the construction and operation of the Virgo detector and the creation and support of the EGO consortium. The authors also gratefully acknowledge research support from these agencies as well as by the Council of Scientific and Industrial Research of India, the Department of Science and Technology, India, the Science & Engineering Research Board (SERB), India, the Ministry of Human Resource Development, India, the Spanish Agencia Estatal de Investigación, the Vicepresidència i Conselleria d'Innovació Recerca i Turisme and the Conselleria d'Educació i Universitat del Govern de les Illes Balears, the Conselleria d'Educació Investigació Cultura i Esport de la Generalitat Valenciana, the National Science Centre of Poland, the Swiss National Science Foundation (SNSF), the Russian Foundation for Basic Research, the Russian Science Foundation, the European Commission, the European Regional Development Funds (ERDF), the Royal Society, the Scottish Funding Council, the Scottish Universities Physics Alliance, the Hungarian Scientific Research Fund (OTKA), the Lyon Institute of Origins (LIO), the Paris Île-de-France Region, the National Research, Development and Innovation Office Hungary (NKFIH), the National Research Foundation of Korea, Industry Canada and the Province of Ontario through the Ministry of Economic Development and Innovation, the Natural Science and Engineering Research Council Canada, the Canadian Institute for Advanced Research, the Brazilian Ministry of Science, Technology, Innovations, and Communications, the International Center for Theoretical Physics South American Institute for Fundamental Research (ICTP-SAIFR), the Research Grants Council of Hong Kong, the National Natural Science Foundation of China (NSFC), the Leverhulme Trust, the Research Corporation, the Ministry of Science and Technology (*MOST*), Taiwan and the Kavli Foundation. The authors gratefully acknowledge the support of the NSF, STFC, INFN and CNRS for provision of computational resources. This work has been assigned LIGO document number LIGO-P190425.

*Software:* The detection of the signal and subsequent significance evalution have been performed using the GSTLAL-based inspiral software pipeline (Cannon et al. 2012; Privitera et al. 2014; Messick et al. 2017; Hanna et al. 2019; Sachdev et al. 2019). These are built on the LALSUITE software library (LIGO Scientific Collaboration 2018). The signal was also verified using the PYCBC (Usman et al. 2016; Nitz et al. 2018, 2019), MBTAONLINE (Adams et al. 2016) and SPIIR (Hooper et al. 2012; Luan et al. 2012; Chu 2017; Guo et al. 2018) packages. The parameter estimation was performed with the LALINFERENCE (Veitch et al. 2015) and LALSIMULATION libraries within LALSUITE (LIGO Scientific Collaboration 2018); additional checks were performed using the BILBY library (Ashton et al. 2019) and the DYNESTY Nested Sampling package (Speagle 2019). The estimates of the noise spectra and the postmerger analysis were performed using BAYESWAVE (Cornish & Littenberg 2015; Littenberg & Cornish 2015). The sky map plot has made use of Astropy,[206] a community-developed core Python package for Astronomy (Astropy Collaboration et al. 2013; Price-Whelan et al. 2018) and `ligo.skymap`.[207] All plots have been prepared using Matplotlib (Hunter 2007).

## Appendix

These Appendices provide more details from the analysis of GW190425. In Appendix A we explicitly give the data channels used for the analyses. In Appendix B we give more details of the triggers produced using multiple search pipelines and describe a single-detector-trigger background analysis using the PYCBC pipeline. In Appendices C and D, we give more details on the source properties and studies of the differences resulting from the use of different waveform families. In Appendix E, we discuss what we can learn about NS matter and the EoS.

## Appendix A
## Data

The data used in the analysis are described briefly in Section 4. For completeness, here we also give the channel names within the gravitational-wave frame format (LIGO Scientific Collaboration & Virgo Collaboration 2019c) files containing the data that we used. We used data from LLO that had been cleaned to remove lines from calibration and from known environmental artifacts (Davis et al. 2019; Driggers et al. 2019), which was stored in the channel name `L1:DCS-CALIB_STRAIN_CLEAN_C01`. In the case of Virgo, we used the low-latency data held in the channel name `V1:Hrec_hoft_16384Hz`. Following the removal of noise transients from the data, we created new frame files containing channels named `L1:DCS-CALIB_STRAIN_CLEAN_C01_T1700406_v3` and `V1:Hrec_hoft_16384Hz_T1700406_v3` for LLO and Virgo, respectively. These were the data we used for estimation of the source properties, and they can be found in Abbott et al. (2019b).

## Appendix B
## Detection

In Section 3, we describe the low-latency detection of GW190425 by the GSTLAL matched-filtering pipeline. Here, we discuss the consistency between results from different matched-filtering searches. As discussed in Section 3, consistent triggers had been produced by other low-latency matched-filtering searches, PYCBC LIVE, MBTAONLINE, and SPIIR. These trigger parameters are listed in Table 2. The difference in S/N between the pipelines is due to the different template banks (Sathyaprakash & Dhurandhar 1991; Owen & Sathyaprakash 1999; Harry et al. 2009) and different methods to estimate the noise power spectral density employed by different searches. Each search pipeline also uses a different definition for calculating the signal-consistency-test values, but for all searches, these values are distributed around 1.0 for signals. Chirp mass is a well measured parameter for relatively low-mass systems (see, e.g., Berry et al. 2016), therefore all of the pipelines obtain consistent chirp mass estimates. To verify that the noise transient described in Section 2 has a negligible effect on the detection of GW190425, we applied a window function to zero out the data around the transient (Usman et al. 2016; Abbott et al. 2017b), which resulted in no significant change in the S/Ns and signal-consistency-test values.

[206] http://www.astropy.org
[207] https://lscsoft.docs.ligo.org/ligo.skymap

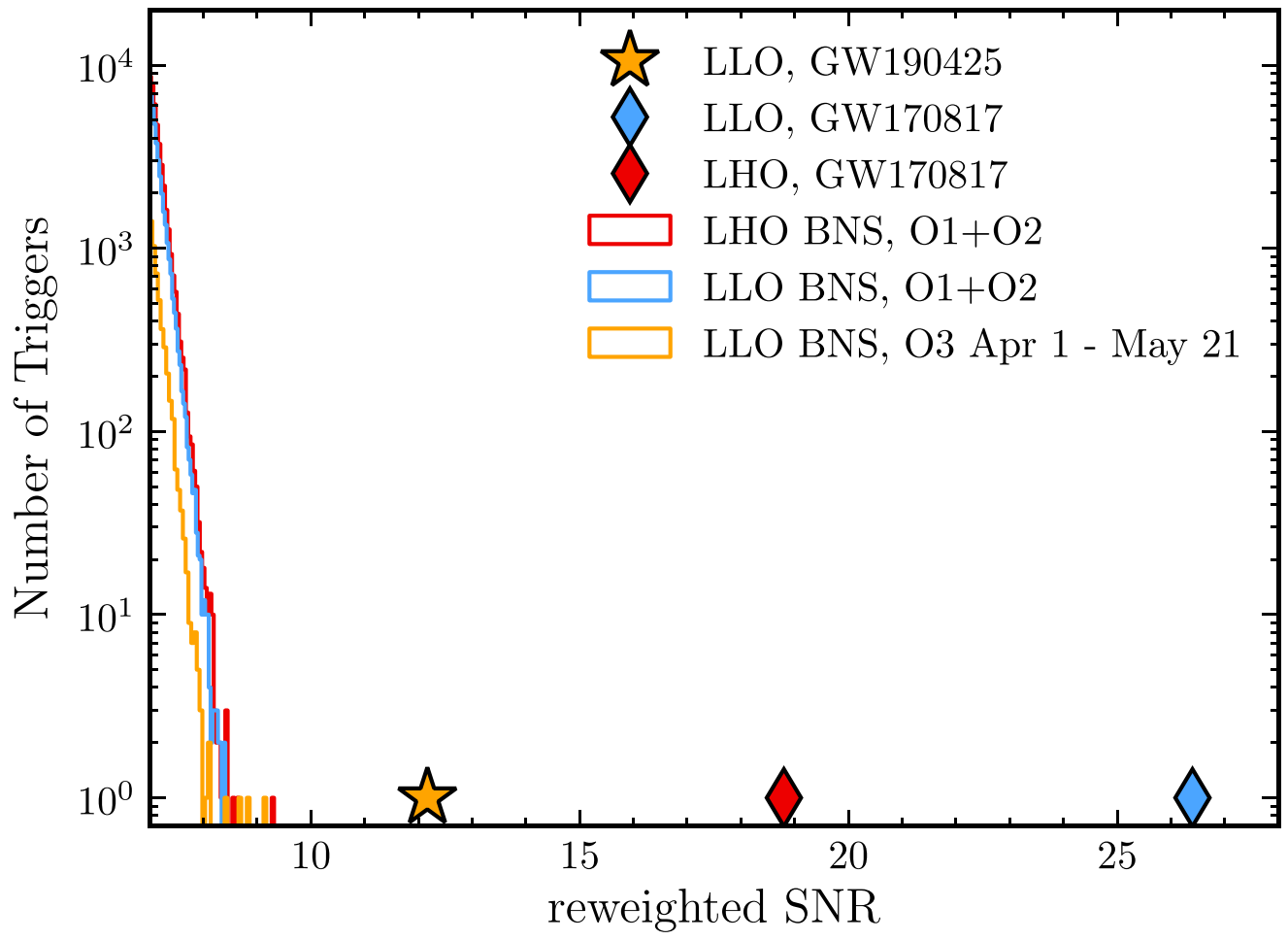

**Figure 6.** Histogram of reweighted S/N for single-detector triggers in the BNS region: the red curve shows the histogram for LLO from O1 and O2, the blue curve shows the histogram for LHO from O1 and O2, and the yellow curve shows the histogram for LLO from the first 50 days of O3. GW190425 is shown as a gold star. It is louder than all the background events. For reference, the LLO and LHO triggers for GW170817 are also shown as blue and red diamonds.

**Table 2**
S/N, Signal-consistency-test Value, and Chirp Mass for GW190425 from Different Low-latency Matched-filtering Pipelines

| Search Pipeline | S/N | Signal-consistency-Test Value | Detector-frame Chirp Mass $M_\odot$ |
|---|---|---|---|
| GstLAL | 12.9 | 0.82 | 1.487 |
| PyCBC Live | 12.1 | 1.03 | 1.487 |
| MBTAOnline | 12.9 | 1.31 | 1.487 |
| SPIIR | 12.0 | 0.79 | 1.487 |

GW190425 was detected as a single-detector event in LLO. As discussed in Section 3, estimating the significance of single-detector candidates is challenging, therefore GW190425 was also compared against background from O1, O2, and the first 50 days of O3 in the BNS region of the parameter space (defined as the parameter space with component masses between 1 and 3 $M_\odot$), shown in Figure 1. Here, we present these results from the PyCBC search, obtained with a template bank constructed using a hybrid geometric-random algorithm, as outlined in Roy et al. (2017, 2019). Histograms of reweighted S/N (Babak et al. 2013; Abbott et al. 2016c), which is the detection statistic for the PyCBC search and a function of S/N and reduced $\chi^2$ (Allen 2005), are shown for LLO and LHO in Figure 6. For reference, the LLO and LHO triggers for GW170817 are shown as blue and red diamonds, respectively. GW190425 is shown as a gold star. It is louder than all the background events.

## Appendix C
## Source Properties

In Section 4, we show posterior probability distributions for the source component masses, system total mass, sky location,

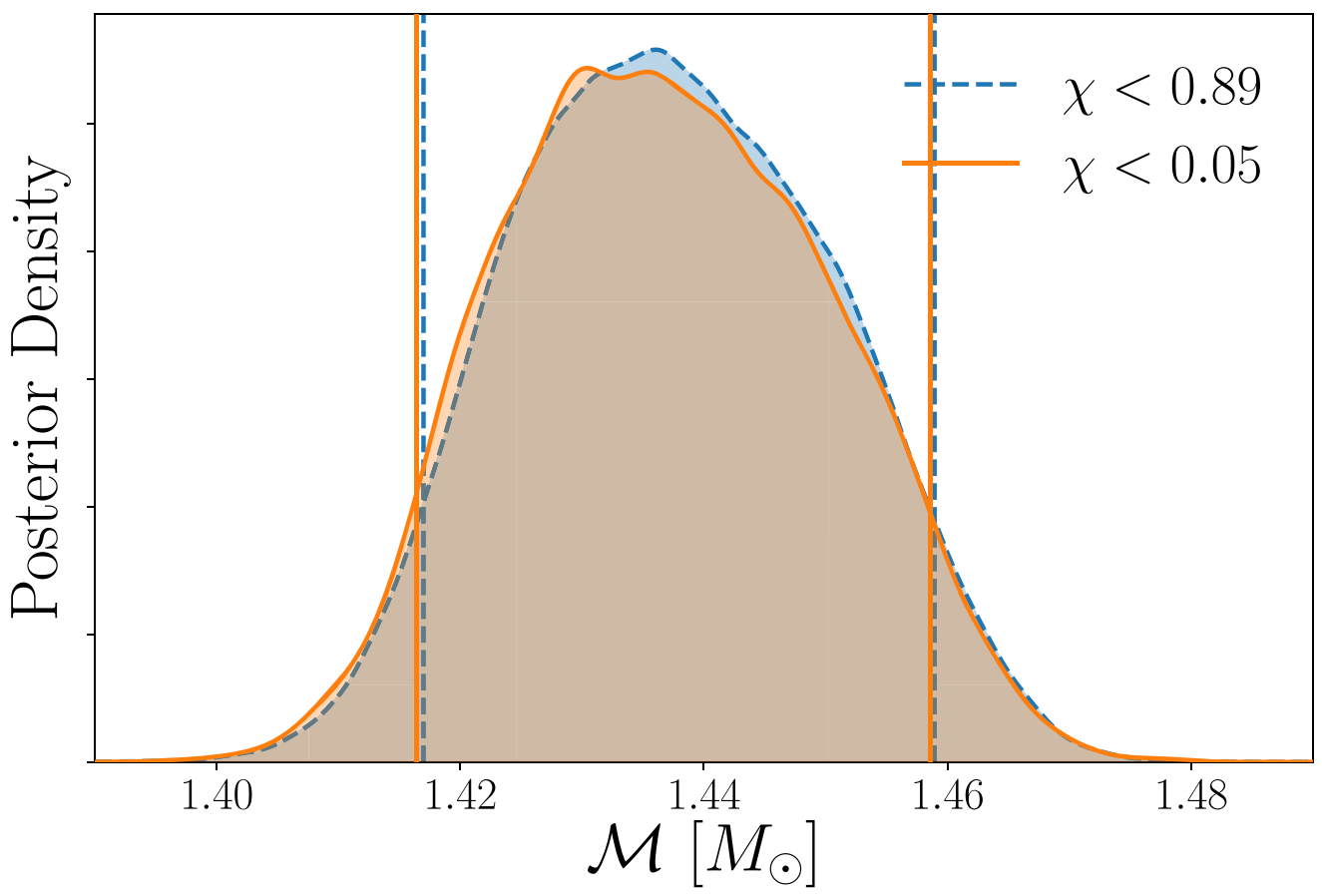

**Figure 7.** Posterior distribution of the source-frame chirp mass for the low-spin prior ($\chi < 0.05$; orange) and high-spin prior ($\chi < 0.89$; blue) analyses using the PhenomPv2NRT waveform. Vertical lines mark the 90% credible interval.

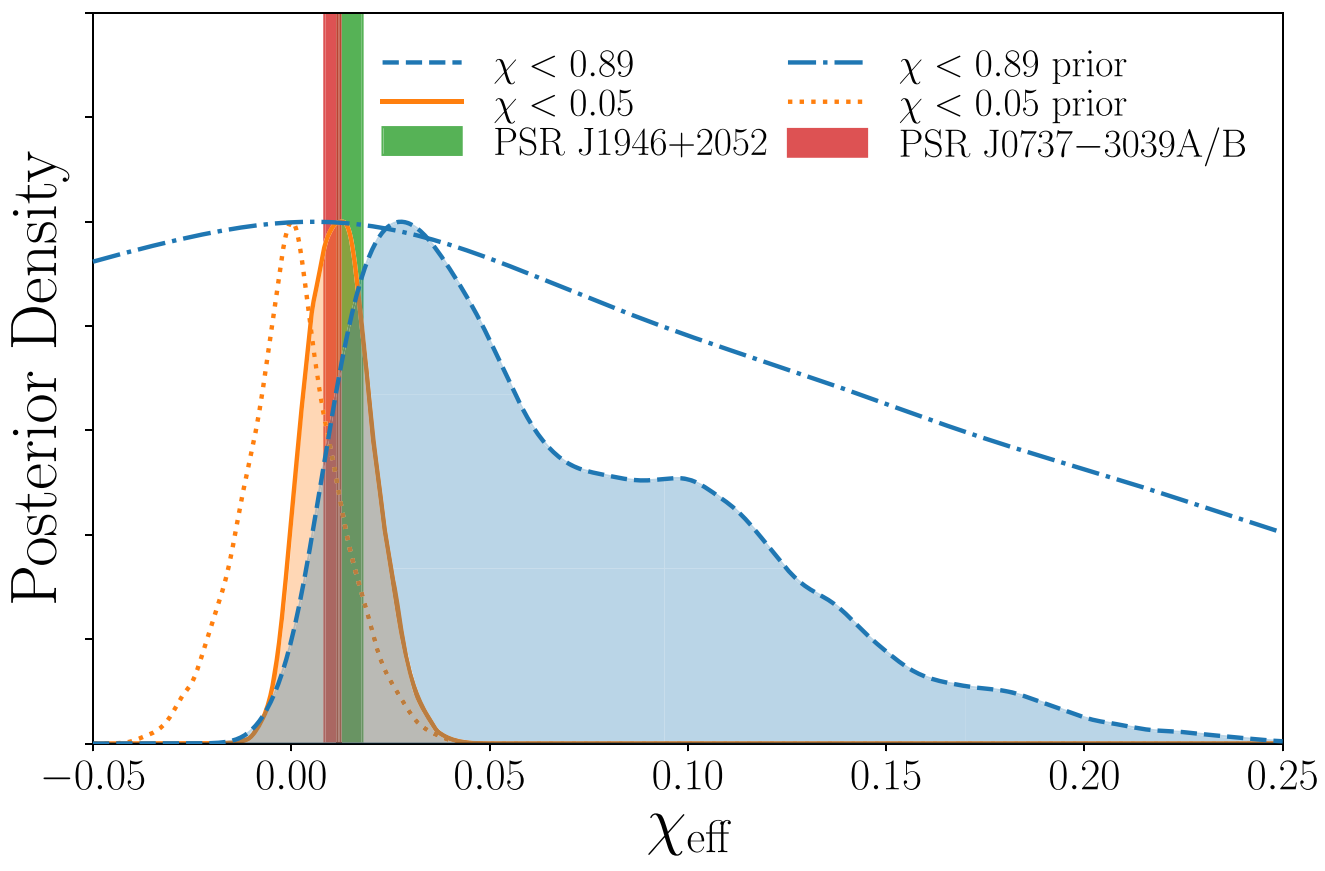

**Figure 8.** Posterior distribution of the effective spin $\chi_{\mathrm{eff}}$ for the low-spin prior ($\chi < 0.05$; orange) and high-spin prior ($\chi < 0.89$; blue) analyses using the PhenomPv2NRT waveform. The prior distribution is shown as the dashed line for each analysis. Predicted at-merger effective spins for two Galactic BNSs, PSR J1946+2052 (green) and PSR J0737−3039 (red), are shown for comparison. The widths of the vertical bands correspond to the uncertainty in $\chi_{\mathrm{eff}}$ due to the unknown NS equation of state, as well as the unknown mass ratio for PSR J1946+2052, as explained in Appendix F.3.

and effective spin parameter $\chi_{\mathrm{eff}}$. These are produced using the PhenomPv2NRT waveform model (Dietrich et al. 2019) and are shown given two different prior assumptions about the component spins: the low-spin case with a uniform prior distribution over $0 \leqslant \chi \leqslant 0.05$, and the high-spin case with a uniform prior distribution over $0 \leqslant \chi \leqslant 0.89$. Here, we provide additional posterior distributions from the analysis using this waveform model.

In Figure 7, we show the source-frame chirp mass. The posteriors are consistent independent of the two different prior assumptions; this is expected as the chirp mass is a particularly well measured property of the signal (see, e.g., Poisson & Will 1995; Berry et al. 2015; Farr et al. 2016).

In Figure 8, we show the posterior distribution of the mass-weighted linear combination of the spins, known as the effective spin parameter $\chi_{\mathrm{eff}}$ (see, e.g., Equation (3) of Abbott et al. 2019d, and associated references) alongside the prior distribution. For the low-spin case, the effective spin posterior is dominated by the informative prior (i.e., the 0.05 upper bound on spin magnitude). For the high-spin prior, which is close to flat over the range of interest, the value of $\chi_{\rm eff} = 0$ is excluded from the 90% posterior credible interval, with 98.8% probability for $\chi_{\rm eff} > 0$. However, as discussed in the Section 4, a comparison of analyses both with and without spin effects present showed no evidence for a spinning system being favored over a non-spinning one. Figure 8 also shows estimates of $\chi_{\rm eff}$ for the two highest-spin Galactic BNS systems that are expected to merge within a Hubble time, showing that the effective spin of the GW190425 source is consistent with these systems.

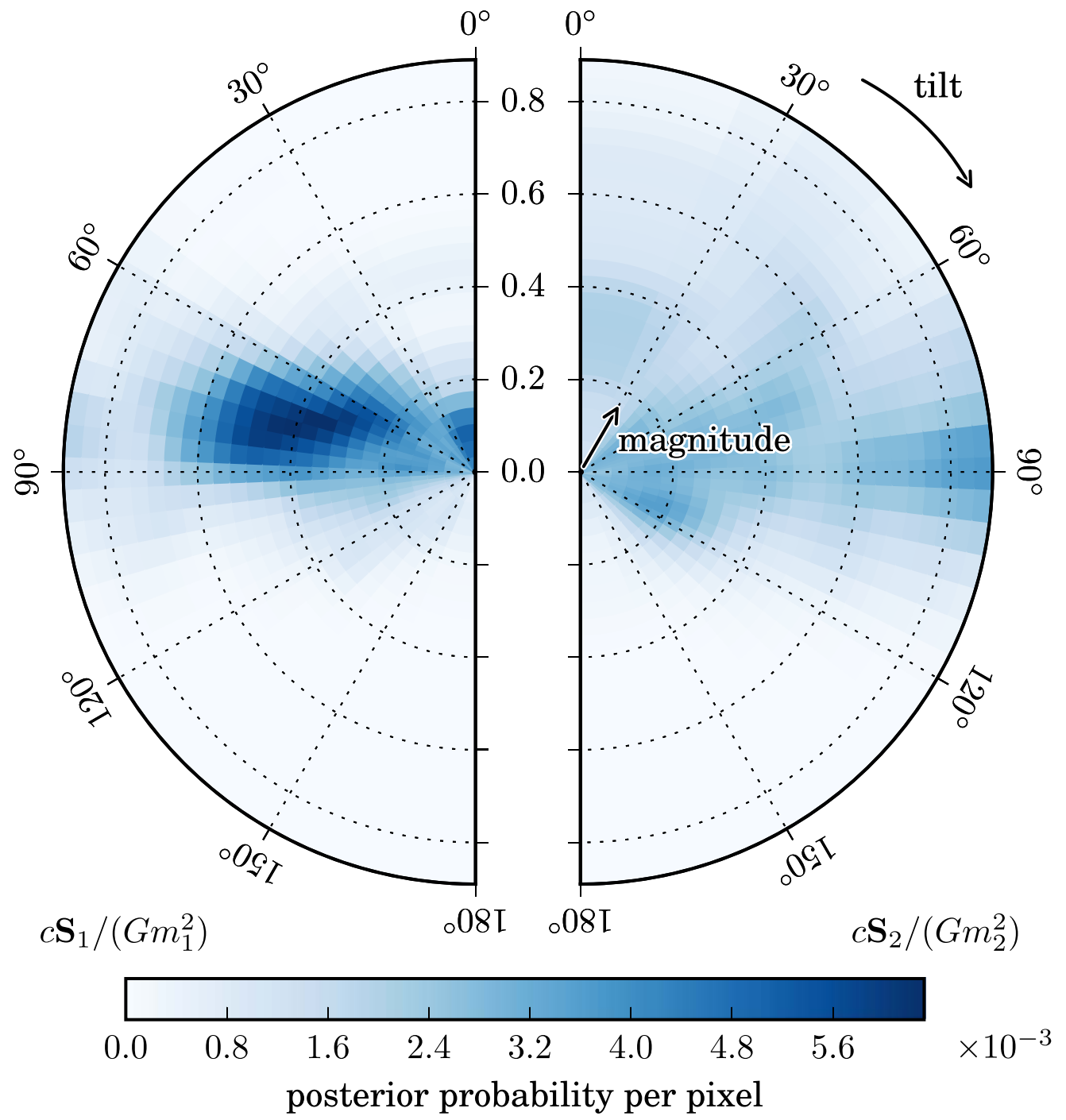

**Figure 9.** Component spin parameter posteriors, marginalized over the azimuthal angle and plotted with respect to the orbital angular momentum. This is shown for the high-spin prior using the PhenomPv2NRT waveform at a reference frequency of 20 Hz. A tilt angle of 0° indicates alignment with the orbital angular momentum.

The precessing spin model PhenomPv2NRT allows one to probe the spin-induced precession of the binary. In Figure 9, we plot the inferred component spin magnitudes and orientations from the high-spin prior results. We are able to rule out modest anti-aligned spin. Degeneracy between aligned spin and mass ratio makes it difficult to measure aligned spin. In Figure 10, we show the posterior distribution for the effective precession parameter $\chi_{\rm p}$ (Schmidt et al. 2015) along with the prior. This illustrates that the data are largely uninformative about the precession of GW190425, with the posterior showing only slight differences compared to the prior.

The luminosity distance $D_{\rm L}$ to the source of GW190425 is given in Table 1 as $159^{+69}_{-72}$ Mpc and $159^{+69}_{-71}$ Mpc for the low- and high-spin priors, respectively. The spin prior does not have a significant effect on the estimation of the distance. The distance is, however, strongly correlated with the angle of the total angular moment vector with respect to the line of sight, $\theta_{JN}$ (see, e.g., Cutler & Flanagan 1994; Nissanke et al. 2010; Abbott et al. 2016d). In Figure 11, we show the marginalized posterior distributions for $D_{\rm L}$ and $\theta_{JN}$. This demonstrates that for GW190425, the inclination angle is not a well measured property.

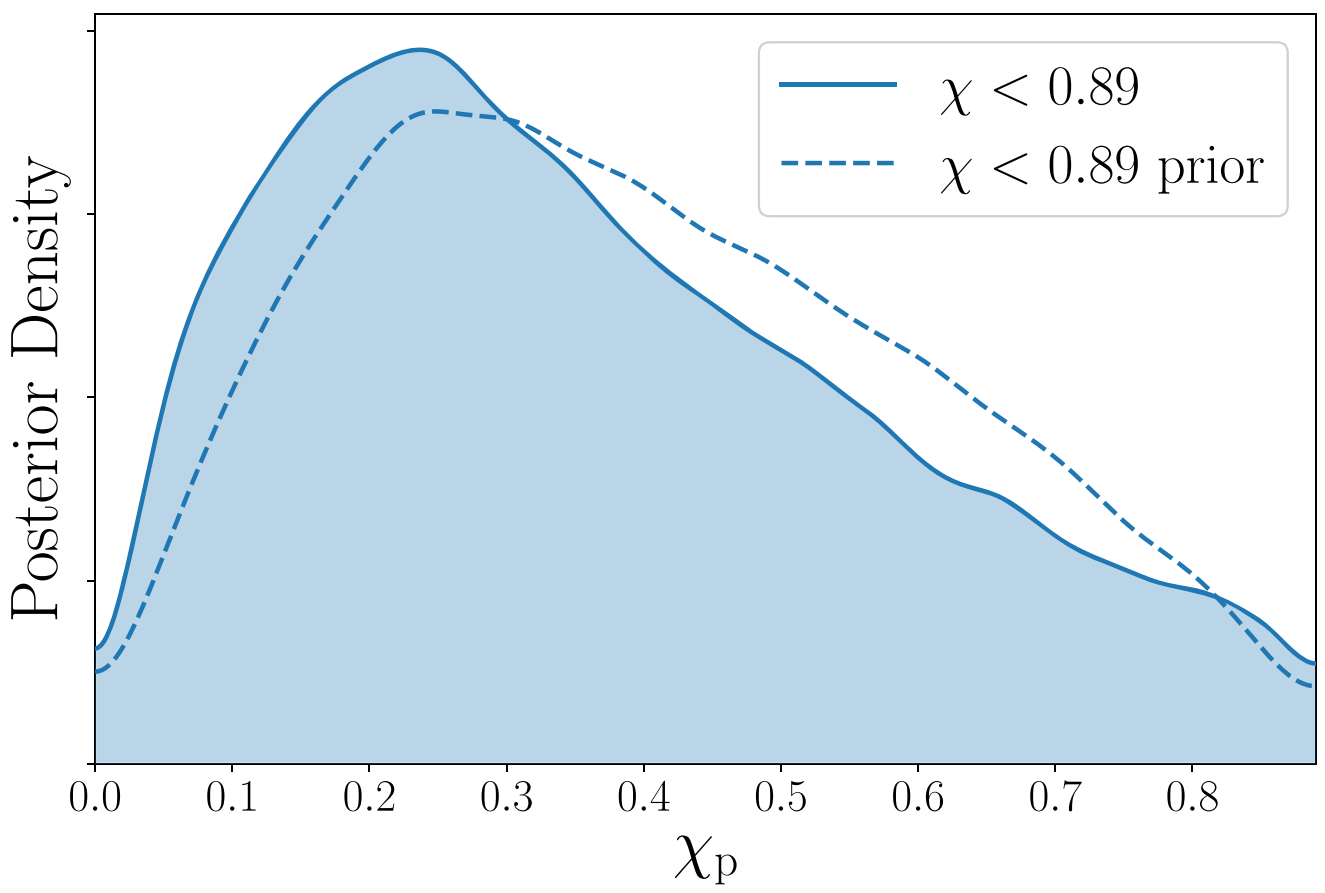

**Figure 10.** $\chi_{\rm p}$ posterior distribution plotted at a reference frequency of 20 Hz from the results using the PhenomPv2NRT waveform. The prior distribution is shown as the dashed line.

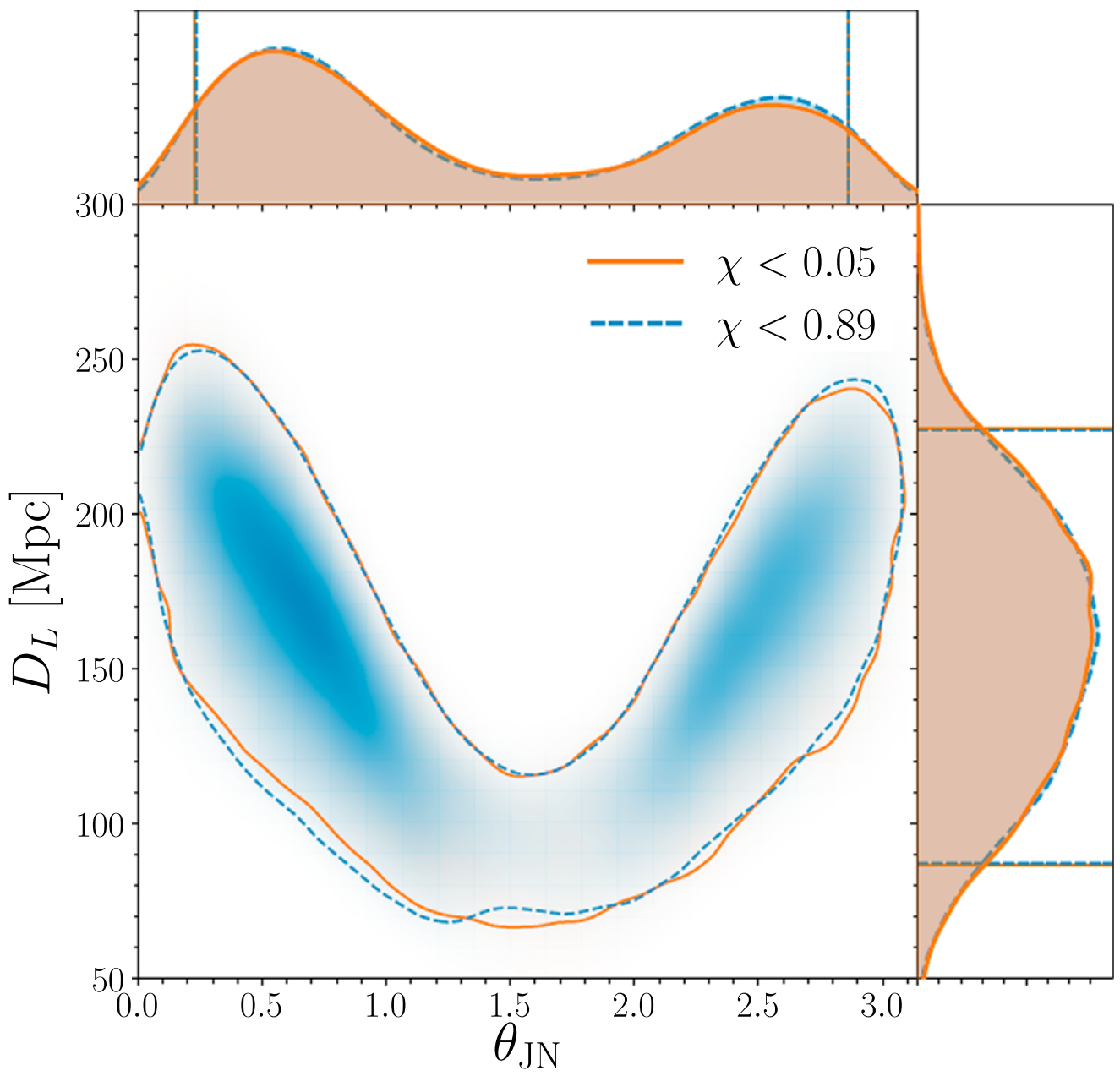

**Figure 11.** Joint posterior distribution of the luminosity distance and $\theta_{JN}$ from the results using the PhenomPv2NRT waveform. Lines in the two-dimensional plot mark the 90% credible interval. Vertical and horizontal lines in the one-dimensional plots mark the extent of the 90% credible interval.

## Appendix D
## Source Properties: Waveform Systematics

The results discussed in Section 4 were obtained with PhenomPv2NRT, a waveform model that augments the numerical relativity (NR)-informed phenomenological description of gravitational-wave phase of a spinning precessing point-particle baseline model (Hannam et al. 2014; Khan et al. 2016) with an analytical effective description of the tidal phase (Dietrich et al. 2017), informed by both EOB waveforms obtained with the TEOBResumS model (Damour & Nagar 2010; Vines et al. 2011; Damour et al. 2012; Bernuzzi et al. 2015; Nagar et al. 2018) (up to the late inspiral) and NR simulations for the last few orbits up to merger. The model also includes spin-induced quadrupole moments (Poisson 1998; Bohé et al. 2015). As PhenomPv2NRT is the only model that includes both spin-precession and tidal effects, it represents a good compromise between physical accuracy and computational efficiency.

Tests of PhenomPv2NRT, EOB tidal models, and NR simulations on simulated signals have suggested that the different models can result in inconsistencies in the inferred BNS parameters for higher $S/N \sim 100$ (Dudi et al. 2018; Messina et al. 2019). These comparisons were restricted to low spins (Dietrich et al. 2019), and comprehensive studies of waveform systematics (including precessing effects) for large spins have not yet been carried out (Messina et al. 2019). The effort to test waveform models for BNS systems including large, and possibly precessing, spins is made more challenging by the lack of NR simulations of appropriate length. However, since the GW190425 progenitor system may not belong to the same population as Galactic NS binaries, it is of astrophysical interest to also analyze the data of GW190425 with a broad spin prior.

In order to study the effect of waveform systematics, we have obtained results for GW190425 with different waveform approximants, namely SEOBNRv4Tsurrogate (Hinderer et al. 2016; Steinhoff et al. 2016; Bohé et al. 2017; Lackey et al. 2019), IMRPhenomDNRT (Husa et al. 2016; Khan et al. 2016; Dietrich et al. 2017, 2019), and TaylorF2 (Sathyaprakash & Dhurandhar 1991; Poisson 1998; Mikóczi et al. 2005; Arun et al. 2009; Bohé et al. 2013, 2015; Mishra et al. 2016). PhenomDNRT is the spin-aligned version of PhenomPv2NRT and does not include the EOS-dependent spin-quadrupole terms (Poisson 1998; Bohé et al. 2015). TaylorF2 is the standard PN spin-aligned approximant taken at 3.5PN order, which is known to potentially induce biases in the recovery of tidal parameters for large $S/N$, because of the incomplete description of the point-mass phasing (Messina et al. 2019). SEOBNRv4Tsurrogate is the faster, frequency-domain surrogate of the tidal, spin-aligned effective-one-body model SEOBNRv4T (Hinderer et al. 2016; Steinhoff et al. 2016), obtained using Gaussian process regression. The SEOBNRv4Tsurrogate analyses are under investigation, but were not fully converged at the time of writing; the preliminary results show qualitatively consistent masses and spins with the other waveforms. Nonetheless, our investigations indicate that, given the moderate $S/N$ of GW190425, our findings are robust with respect to waveform systematics in the spin-aligned dynamics for both high and low spins. The spin-precession dynamics in PhenomPv2NRT have only been verified for signals of shorter length (Dietrich et al. 2019) as no other spin-precessing and tidal model or NR simulation exists to date.

In Figures 12(a)–(f), we reproduce the figures from Section 4 and Figure 11, but comparing the three waveforms for the high- and low-spin cases. In Figure 12(a), for the high-spin, component mass plot, the reduced-order quadrature method (Smith et al. 2016) imposes an arbitrary lower bound on the component mass of $1\,M_\odot$ for PhenomDNRT (Smith 2019) and PhenomPv2NRT (Baylor et al. 2019). To use the reduced-order quadrature basis, we applied an additional constraint on the detector-frame chirp mass to be between $1.485\,M_\odot$ and $1.490\,M_\odot$ for the PhenomPv2NRT model and between $1.42\,M_\odot$ and $2.6\,M_\odot$ for the PhenomDNRT model. The posterior distribution for both the PhenomPv2NRT and PhenomDNRT models shows support at the lower detector frame component mass bound, see Figure 13. However, for TaylorF2, we reduced the lower bound to $0.7\,M_\odot$ after initial results demonstrated a similar feature; we also used a prior on both component $\Lambda$ values with a maximum of 5000, unlike those for the IMRPhenom waveforms. For TaylorF2, the secondary component mass posterior falls off well above $0.7\,M_\odot$, but otherwise remains broadly consistent with PhenomDNRT above the cut. Under the assumption that results from PhenomDNRT with an unconstrained mass prior would fall off below the cut with similar overall slope to the other waveforms, we obtain consistent constraints on the component mass ranges using the three different waveform models. For the remaining figures, the results are also broadly consistent, with some approximants showing more conservative limits on the tidal parameters, providing evidence that our GW190425 conclusions are not subject to systematic bias due to the choice of waveforms. This is consistent with conclusions from the similar systematics study performed on GW170817 (Abbott et al. 2019d), especially given the lower $S/N$ of GW190425.

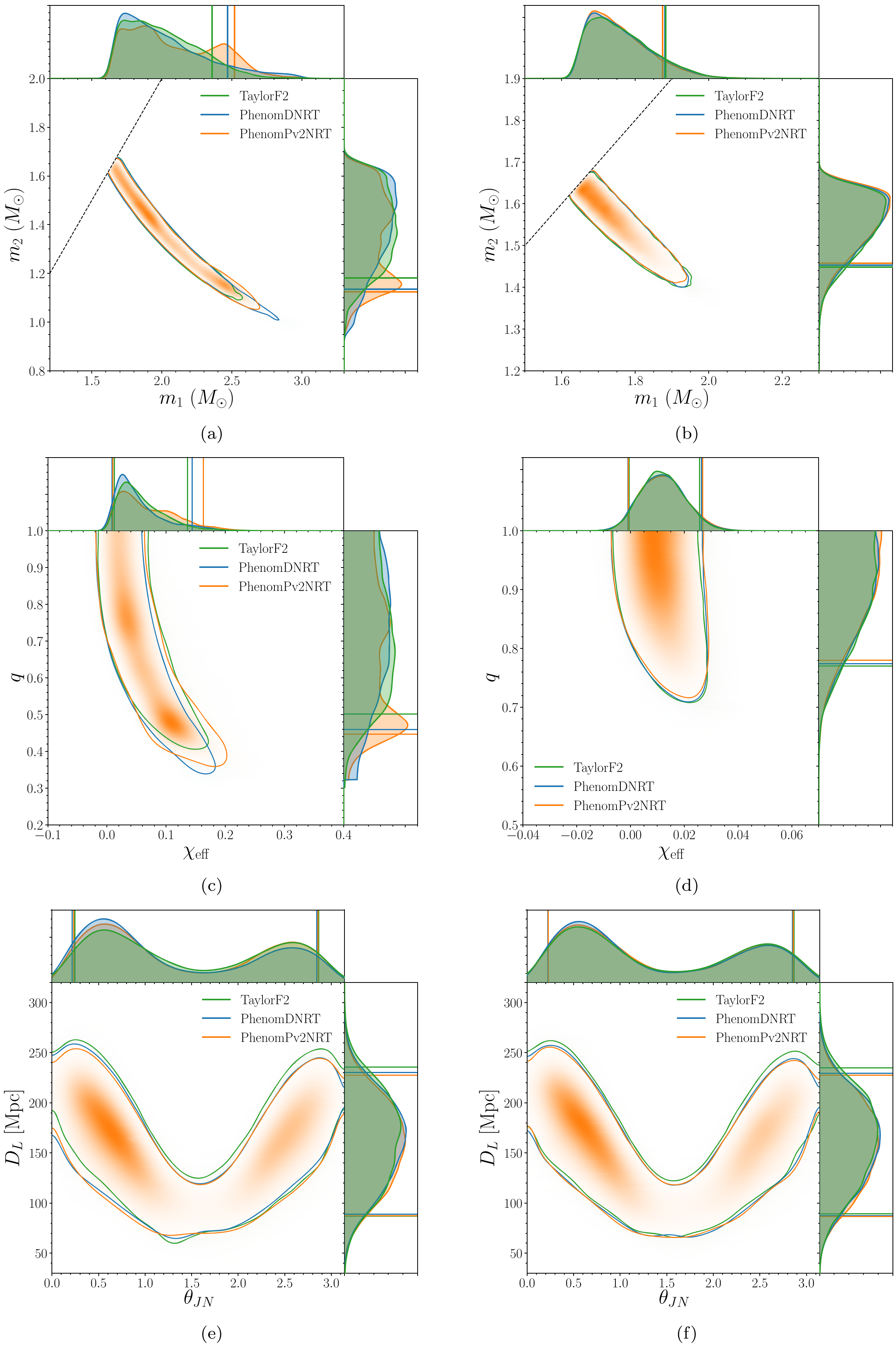

**Figure 12.** Checks of the robustness to waveform systematics for GW190425. Posterior distributions are given for the component masses, mass-ratio and effective spin, and distance and inclination for the high-spin prior (left-hand side) and low-spin prior (right-hand side). Vertical and horizontal lines in the one-dimensional plots mark the extent of the 90% credible interval.

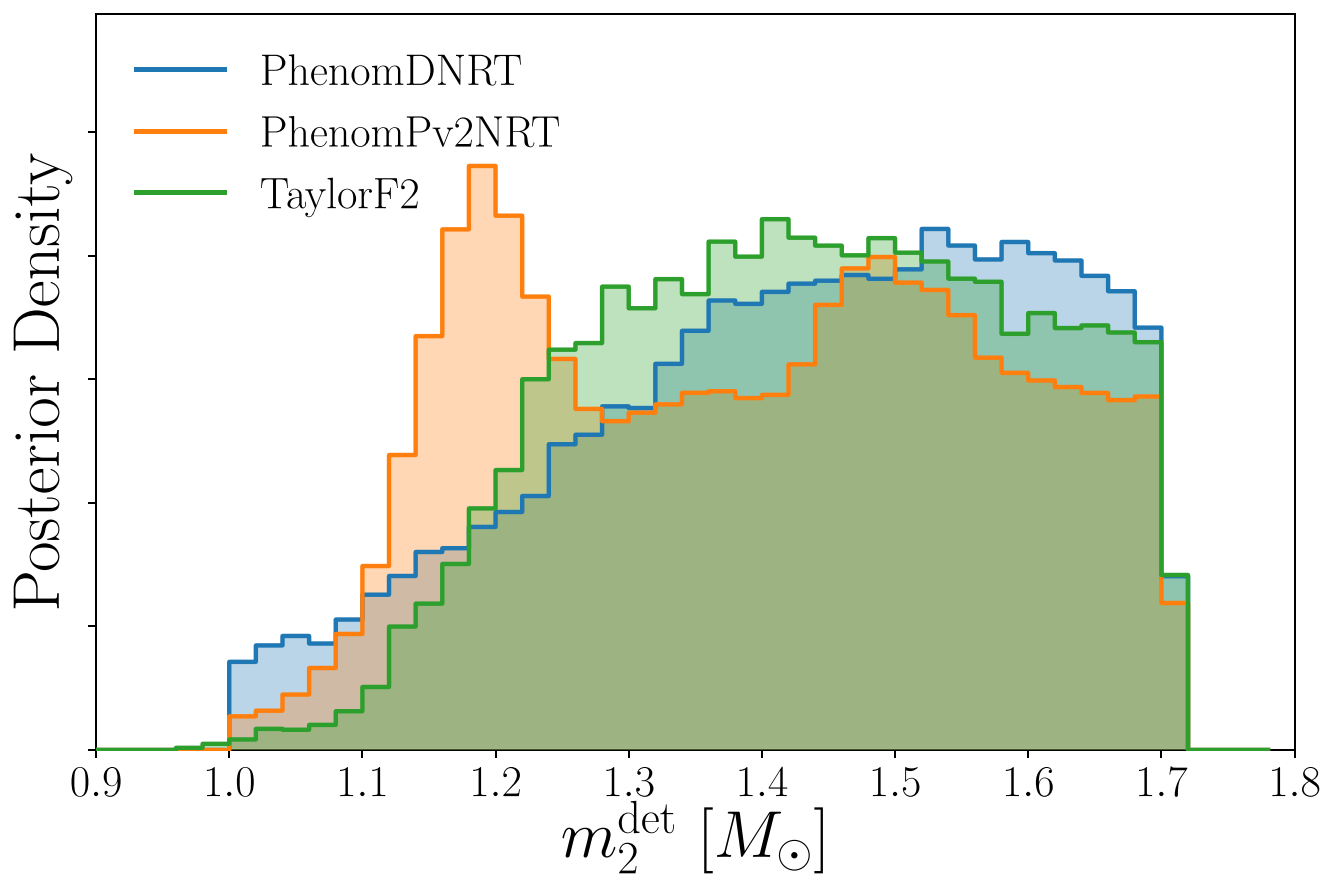

**Figure 13.** High-spin posterior distributions for the detector-frame secondary component mass. For PhenomDNRT and PhenomPv2NRT, an arbitrary low prior bound of 1 $M_\odot$ is imposed by the reduced-order quadrature bases used for the analysis. For TaylorF2, we instead apply a lower bound of 0.7 $M_\odot$.

## Appendix E
## NS Matter Effects

The macroscopic properties of a NS are related to its microphysics. The EOS, i.e., the relation that links the star's internal pressure to its density, can be mapped to a relationship between macroscopic observables of NSs such as their mass, radius, moment of inertia, and tidal deformability (Lattimer & Prakash 2001). For a coalescing BNS system, this information is primarily encoded in the mass-weighted tidal parameter $\tilde{\Lambda}$ (Flanagan & Hinderer 2008) which generates the leading-order tidal effects in the waveforms described previously. The smaller quadrupole–monopole effect (Poisson 1998), which is quadratic in the spins and caused by the deformations induced by rotation, is included in our models by means of approximate universal relations with the tidal deformability of NSs (Yagi & Yunes 2017). All analyses presented in the following paragraphs assume that both components of the GW190425 coalescence were NSs.

### F.1. Tidal Constraints

From the analyses detailed in the main text, the mass-weighted tidal parameter $\tilde{\Lambda}$ is constrained to 1200 and 1900 for low- and high-spin priors, respectively. Such upper limits are obtained by employing a prior that is uniform in the component's tidal deformability. This choice leads to a prior distribution for $\tilde{\Lambda}$ that correlates with the mass ratio: any information about $q$ inferred from the observational data will have an impact on the posterior of $\tilde{\Lambda}$, even if the signal does not carry any information on the tidal deformabilities, thus complicating the interpretation of the results (see also Kastaun & Ohme 2019). We therefore construct a second prior, flat in $\tilde{\Lambda}$ at any mass ratio, which avoids the correlation with $q$ and does not disfavor low $\tilde{\Lambda}$ values. We employ the standard reweighting procedure that explicitly considers the $q$ dependence of the $\tilde{\Lambda}$ prior $\pi(\tilde{\Lambda})$ before marginalization over $q$. To generate the posterior distribution of $\tilde{\Lambda}$ obtained with a flat prior, without modifying the mass ratio prior $\pi(q)$, we replace the original prior $\pi_{\rm old}(\tilde{\Lambda}, q) = \pi_{\rm old}(\tilde{\Lambda}|q)\, \pi(q)$ produced by uniform component sampling with a prior $\pi_{\rm new}(\tilde{\Lambda}, q) = \pi_{\rm new}(\tilde{\Lambda})\, \pi(q)$, where $\pi_{\rm new}(\tilde{\Lambda})$ is independent of $q$ and flat over a fixed range contained within the original prior and extending well above the posterior support. We evaluate

$$p_{\rm flat}(\tilde{\Lambda}|d) = \int dq \frac{p_{\rm old}(\tilde{\Lambda}, q|d)\pi_{\rm new}(\tilde{\Lambda}|q)}{\pi_{\rm old}(\tilde{\Lambda}|q)}, \tag{1}$$

and confirm that this procedure reproduces the results of runs sampled uniformly in $\tilde{\Lambda}$. After reweighing the posterior to correspond to a flat prior in $\tilde{\Lambda}$, we find that under the low-spin ($\chi < 0.05$) assumption $\tilde{\Lambda}$ is constrained to $\tilde{\Lambda} \leqslant 650$ for all approximants, whereas the high-spin case ($\chi < 0.89$) leads to larger differences in upper constraints (Figure 14). These values are less constraining than those quoted above, and this ambiguity can be traced back to the fact that tidal information in current gravitational-wave signals is not strong.

For comparison, we combine the mass posterior distributions of GW190425 with EoS samples taken from Abbott et al. (2017b) and compute upper limits for $\tilde{\Lambda}$. We consider sets of samples of GW190425's component masses ($m_1$, $m_2$) and of GW170817's spectral coefficients ($\gamma_0, \ldots, \gamma_3$), each randomly drawn from its respective posterior distribution. The spectral coefficients are then mapped into an EoS through Equation (7) of Lindblom (2010), which is then employed together with the component masses to compute tidal parameters. By additionally interpreting the heavier body as a BH with $\Lambda_1 = 0$ whenever its mass exceeds the maximum mass supported by the EoS, we infer upper limits on $\tilde{\Lambda}$ of 230 and 220 for the GW190425 mass distributions with low- and high-spin priors, respectively. Following this procedure, the EoS constraints from GW170817 are effectiely mapped to the GW190425 progenitor's mass scale. These limits are much tighter than those derived from GW190425 alone.

### F.2. Radii and EoS

To immediately obtain a rough estimate on the bound that we can place on the NS radius, we use the $\tilde{\Lambda}$ upper limit obtained above, the individual mass measurements of GW190425 and the EoS-insensitive relations of Yagi & Yunes (2016, 2017), to find $R < 16$ km and $R < 15$ km with high- and low-spin priors, respectively. Such values suggest that the GW190425 signal is too weak to provide further EoS constraints. We obtain similar results by employing different relations (De et al. 2018; Raithel 2019), which produce results consistent with those obtained by comparing directly with a set of EoS models (e.g., Annala et al. 2018). Under the assumption that both NSs are described by a single fundamental EoS, more detailed analyses can be performed. The quasi-universal relations explored in Yagi & Yunes (2016, 2017) allow the inference of the tidal deformabilities and radii from sampling the mass ratio and the symmetric combination of the tidal parameters (Abbott et al. 2017b; Chatziioannou et al. 2018); the same information can be obtained by direct parameterization of the EoS above densities of $10^{14}\,{\rm g\,cm^{-3}}$ ($10^{17}\,{\rm kg\,m^{-3}}$), and fixing the low-density crust to obey the SLy EoS (Douchin & Haensel 2001) as described in Lackey & Wade (2015), Carney et al. (2018), and Abbott et al. (2017b). This second approach additionally allows for EoS reconstruction and can be supplemented with constraints that incorporate astrophysical observations, e.g., by introducing a term in the likelihood that depends on the maximum mass supported by the NS EoS (Alvarez-Castillo et al. 2016; Miller et al. 2020). While the above approach is preferred, given the uninformative nature of GW190425 on matter effects, and for consistency with Abbott et al. (2017b), we impose a sharp cut on the lower bound of the maximum EoS mass, and reject samples that do not support at least 1.97 $M_\odot$ (Antoniadis et al. 2013). Due to the high inferred mass of the heaviest binary component, results obtained through universal relations could be subject to unexplored modeling systematics, as such relations are primarily applied to lower-mass NSs (Yagi & Yunes 2017). On the other hand, a direct EoS parameterization is expected to more accurately capture the properties of high-mass NSs (Lindblom 2010); given the moderate S/N of the event, it is not expected that any significant systematic issue will arise from spectral analyses. To verify this we compare the component mass distributions to those shown in Figures 12(a) and (b), with corresponding results obtained when using the spectral EoS parameterization and find only minor differences that can be attributed to the lack of tidal information contained in GW190425 and to requiring that the sampled EoS support the sampled component masses. We therefore explicitly show only results obtained with the latter. Figure 15 shows the reconstructed EoS. We compute pressure at twice and six times nuclear density (see Table 3). In Figure 16, instead, we show the marginalized 2D distribution of the masses and radii obtained through our spectral investigations. It shows the results from the high- and low-spin priors and for both cases when the restriction that the EoS support masses of $1.97 M_\odot$ is or is not applied. All cases result in a radius upper constraint of approximately $R < 15$ km at 90% credible level. Both confirm that GW190425 does not carry significant novel information on the NS EoS and our constraints a posteriori are similar to our prior beliefs.

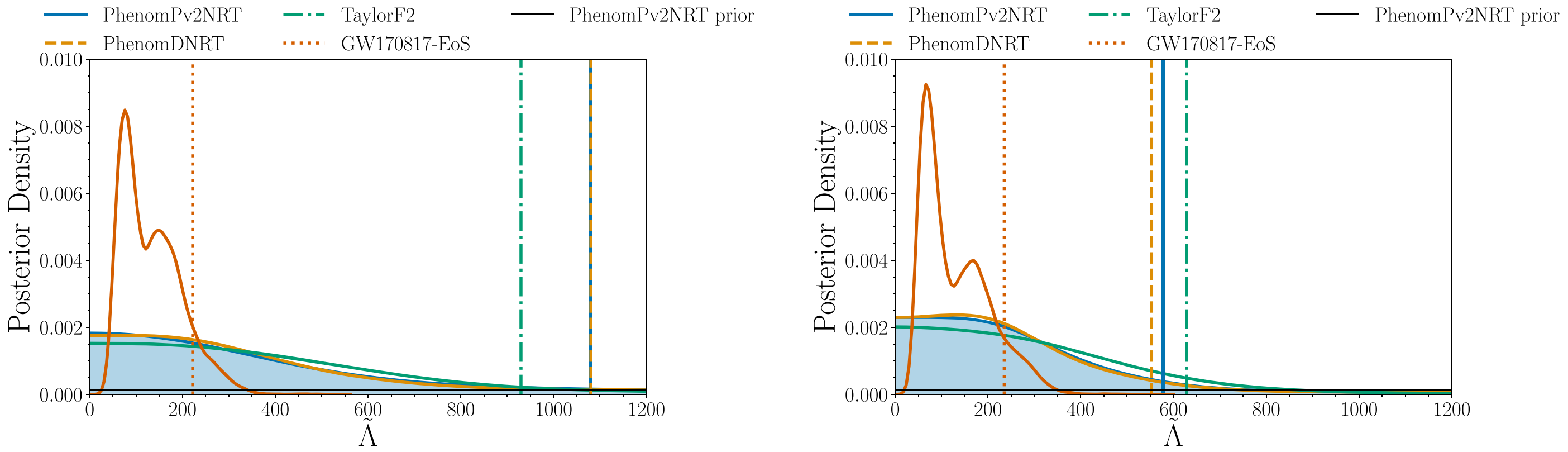

**Figure 14.** Distributions of the reweighted tidal parameter $\tilde{\Lambda}$ for (left) high-spin and (right) low-spin priors, together with their upper 90% one-sided credible interval (vertical lines). Shown in red is the distribution of $\tilde{\Lambda}$ obtained by propagating GW170817's constraints to GW190425's mass regime.

### F.3. Spins

To provide context for the GW190425 spin measurement, we calculated the effective spins $\chi_{\rm eff}$ of the two fastest known Galactic BNSs that will merge within a Hubble time, PSR J1946+2052 (Stovall et al. 2018) and PSR J0737−3039 (Burgay et al. 2003). When comparing the pulsar and gravitational-wave observations, the pulsar spin periods $P$ were converted to dimensionless spins $\chi$ via the moment of inertia, which depends on the unknown NS EoS and the pulsar mass. We inferred the pulsar moments of inertia using mass posteriors from Farrow et al. (2019) and samples from the posterior distribution of the spectral parameterization of the EoS obtained from the analysis of GW170817 (Abbott et al. 2017b) as inputs. For each sample, we calculated the moment of inertia from the EoS and NS mass in the slow-rotation approximation (Hartle 1970). Any uncertainty in the pulsar mass was marginalized over as part of this procedure. The binary pulsar effective spins, with error estimates for the EoS and mass uncertainty, follow from the inferred moments of inertia, binary masses, and spins. We have verified that the effective spins obtained in this way agree with those calculated according to the universal-relation-based method of Landry & Kumar (2018) and Kumar & Landry (2019), which uses the $\Lambda_{1.4}$ posterior from GW170817 (Abbott et al. 2017b) as input.

### F.4. Central Density and Pressure

NSs are known to be exceptional laboratories for studying cold matter at extreme densities. GW190425, given its large chirp mass, suggests that gravitational waves can be used to probe such densities. By combining the GW170817 EoS samples with the GW190425 component mass posterior distributions, we compute the implied central pressure and density distributions (Figure 17). We estimate the matter density in the core of the heavier component to be between three and six times nuclear density and the pressure to between $1 \times 10^{35}$ dyn cm$^{-2}$ and $8 \times 10^{35}$ dyn cm$^{-2}$ ($1 \times 10^{34}$ Pa and $8 \times 10^{34}$ Pa), at the 90% credible interval.

### F.5. Nonlinear Tides

Nonresonant, nonlinear fluid instabilities within NSs ($p$–$g$ instabilities) may impact the gravitational waveform, particularly at low frequencies (Venumadhav et al. 2013; Weinberg et al. 2013; Weinberg 2016; Zhou & Zhang 2017). With the same techniques used to analyze GW170817 (Abbott et al. 2019e), we can constrain a phenomenological model for the $p$–$g$ instability (Essick et al. 2016; Abbott et al. 2019e) with GW190425. Unlike the rest of the analyses presented here, we only analyze GW190425 down to 30 Hz to be consistent with the procedure adopted for GW170817. We also assume a high-spin prior ($\chi < 0.89$) uniform in the component of the spins perpendicular to the orbital plane instead of isotropic spin orientations. This favors larger spins a priori (less constraining than the isotropic spin prior) and corresponds to slightly wider mass posteriors, but does not significantly impact our conclusions. GW190425, by itself, is less informative than GW170817 and is only able to rule out phenomenological $p$–$g$ amplitudes above 1.3 times the 90% upper limit obtained from GW170817 at the same credible level. GW190425 produces Bayes factors between models that include $p$–$g$ effects and those that do not of $\ln B^{pg}_{!pg} = 0.1^{+1.3}_{-0.3}$ with the high-spin prior, similar to GW170817. Again, the data are not informative enough to either detect or disprove the existence of $p$–$g$ instabilities. Combining information from both events does not significantly improve existing constraints and is dominated by GW170817.

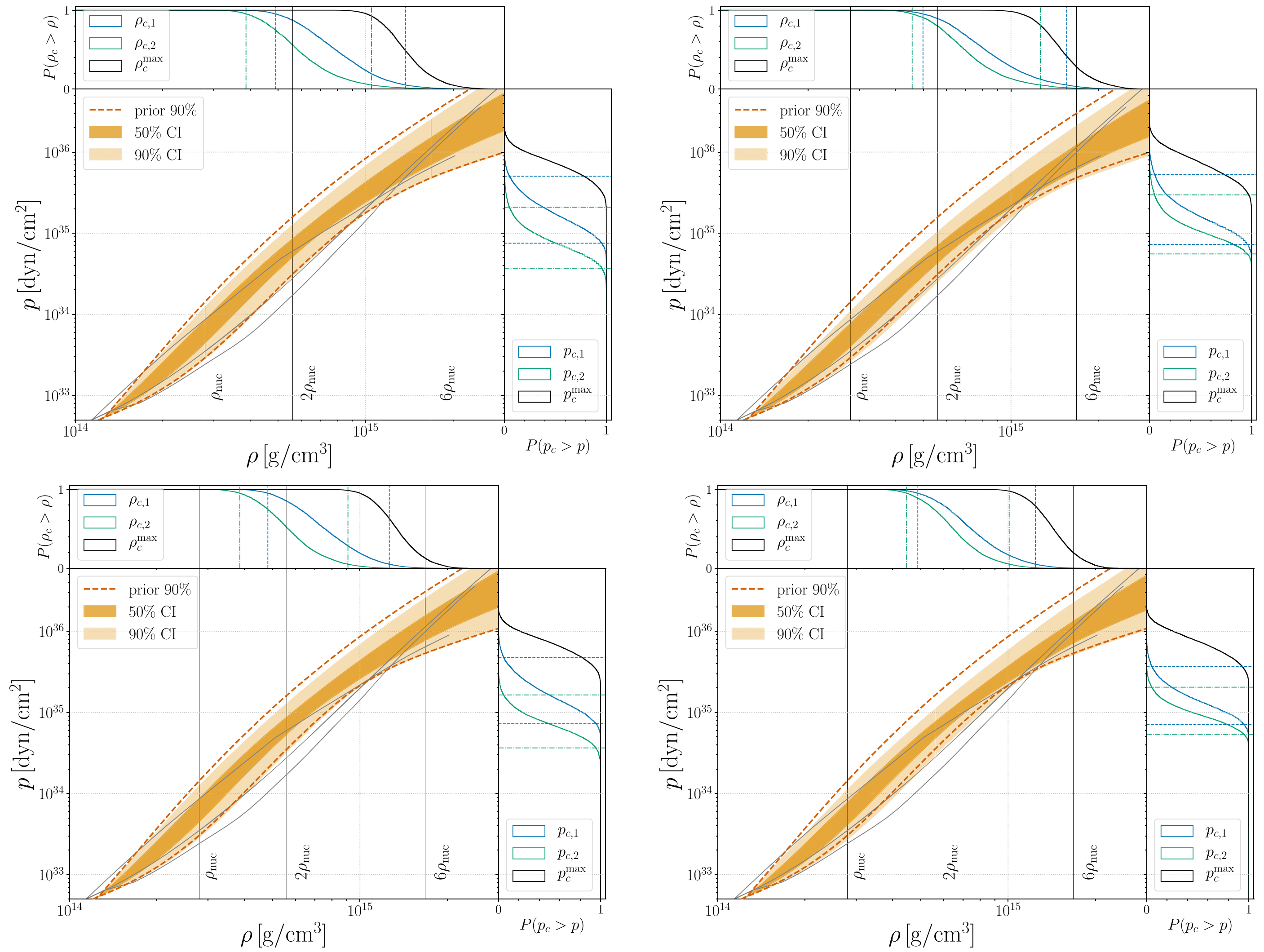

**Figure 15.** 50% and 90% credible levels for the marginalized pressure posteriors $p$ as a function of the rest mass density $\rho$, for (left) high-spin and (right) low-spin priors. Plots displayed in the bottom row are obtained from spectral analyses which additionally require that the EoS support masses above 1.97 $M_\odot$. The prior (dashed red line) is also shown. Vertical lines correspond to one, two, and six times nuclear saturation density. In gray, example EoSs are displayed: from top to bottom at twice nuclear density, H4, APR4, and WFF1. The top and right panels of each figure show, respectively, the central pressure and density cumulative posteriors for the two stars (blue and green) and the heaviest NS supported by the EoS (black). Vertical dashed lines correspond to the 90% credible interval bounds.

**Table 3**
Pressures at Twice and Six Times Nuclear Density and Radii Upper Limits Obtained by Analyzing GW190425 with Spectral Decomposition and Different Choices of Spin and Maximum EoS Mass Priors

| | Low-spin Priors ($\chi < 0.05$) | | High-spin Priors ($\chi < 0.89$) | |
|---|---|---|---|---|
| | Max. Mass | No Max. Mass | Max. Mass | No Max. Mass |
| $p_{2\rho_{nuc}}$ (dyn cm$^{-2}$) $\times 10^{34}$ | $-5.9^{+4.6}_{-3.1}$ | $5.5^{+4.6}_{-3.1}$ | $6.7^{+5.5}_{-3.5}$ | $6.3^{+5.4}_{-3.5}$ |
| $p_{6\rho_{nuc}}$ (dyn cm$^{-2}$) $\times 10^{35}$ | $9.4^{+13.3}_{-4.2}$ | $8.2^{+13.1}_{-3.8}$ | $10.9^{+15.2}_{-5.5}$ | $10.0^{+15.2}_{-5.2}$ |
| $R$ (km) | $\leqslant 14.6$ | $\leqslant 14.4$ | $\leqslant 14.9$ | $\leqslant 14.9$ |

### F.6. Prompt Collapse

Applying the analysis of Agathos et al. (2020) to the total mass of GW190425 allows us to infer the prompt collapse probability and the maximum threshold total mass before merger $M_{thr}$ above which BNSs are expected to promptly collapse into a BH. By relying on the EoS samples of GW170817 (Abbott et al. 2017b) and extrapolating from fits to sparse NR data, obtained from the simulations of Hotokezaka et al. (2011), Bauswein et al. (2013), Zappa et al. (2018), Dietrich et al. (2018), Radice et al. (2018), and Köppel et al. (2019), we estimate the probability of the binary promptly collapsing to be 96% and 97%, for the low- and high-spin priors, respectively. The left panel of Figure 18 shows the distribution of the total mass of the GW190425 system compared to $M_{thr}$. Restricting the EoS samples to consider only those that support a stable nonrotating NS of $1.97\ M_\odot$, leads to an increase of the values of $M_{\rm thr}$, and to updated prompt collapse probabilities of 82% and 88%, for the low- and high-spin priors, respectively. We also compute the behavior of the probability of prompt collapse as a function of the maximum observed NS mass, which is shown on the right panel of Figure 18. Assuming both low- and high-spin priors, the binary is found to have likely undergone prompt collapse.

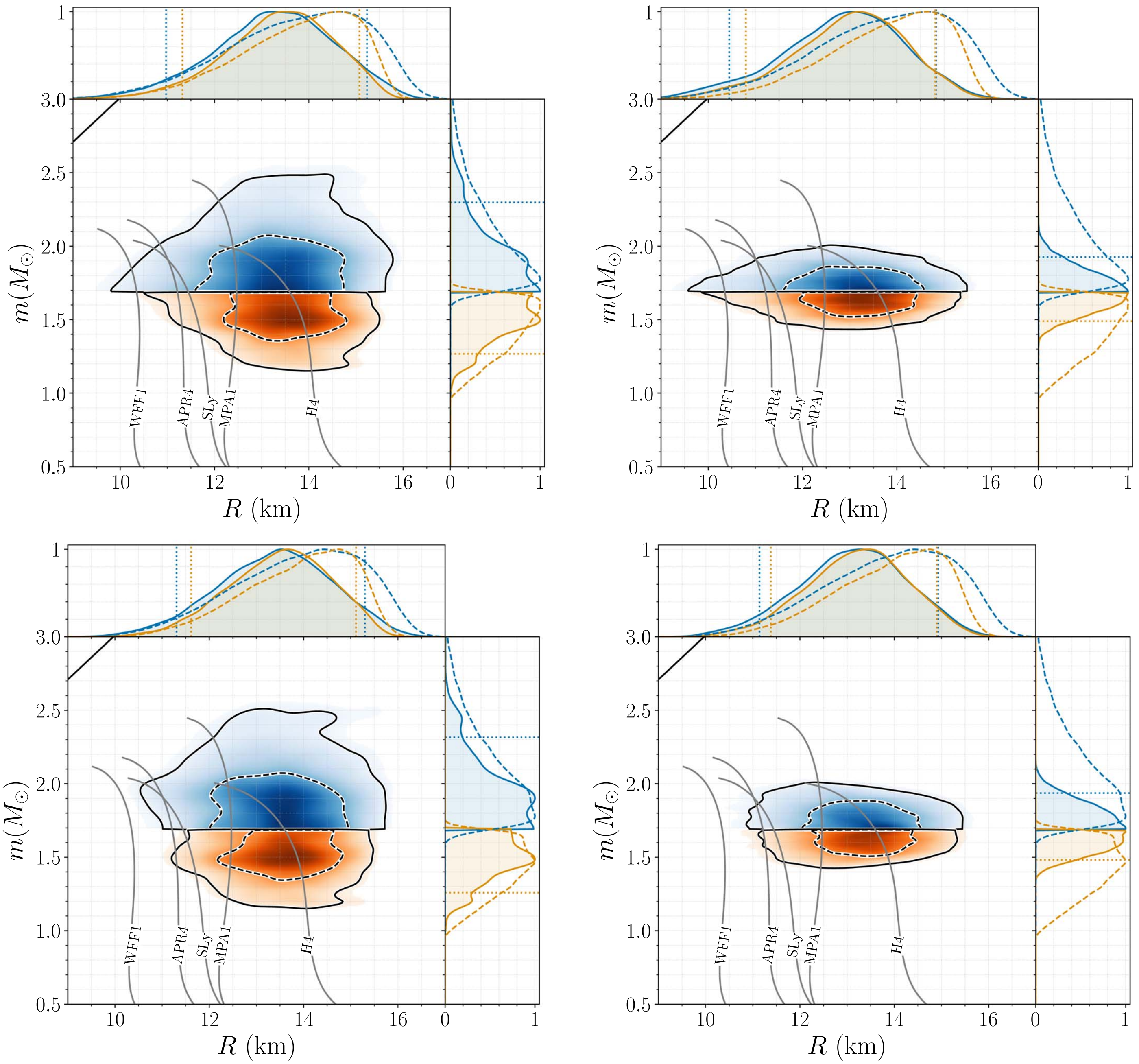

**Figure 16.** Marginalized posterior distributions of the component masses ($m_1$ in blue, $m_2$ in orange) and radii, displayed following the same disposition as Figure 15. Black (black dashed) lines represent 90% (50%) credible limits. Priors of both quantities are shown through dashed lines, while vertical dotted lines indicate the 90% credible intervals. Example mass–radius curves for selected EoSs are overplotted in gray.

### *F.7. Postmerger*

To also consider the unlikely scenario where the remnant object did not promptly collapse, we repeated the high-frequency ($f > 1000$ Hz) unmodeled analysis of Chatziioannou et al. (2017) and Abbott et al. (2019d) to look for any postmerger signal. We used approximately 1 s of strain data around the time of merger. We found no evidence of a statistically significant signal, with a (natural) log Bayes factor of $0.41 \pm 1.13$ in favor of stationary Gaussian noise compared to the signal model. Following Abbott et al. (2017a), we obtained 90% credible upper limits on the strain amplitude spectral density and the energy spectral density of $1.1 \times 10^{-22}\ {\rm Hz}^{-1/2}$ and $0.11\ M_\odot\, c^2\ {\rm Hz}^{-1}$, respectively, for a frequency of 2.5 kHz (see Figure 19). Due to the large distance to the source, these upper limits are less interesting than those obtained from GW170817.

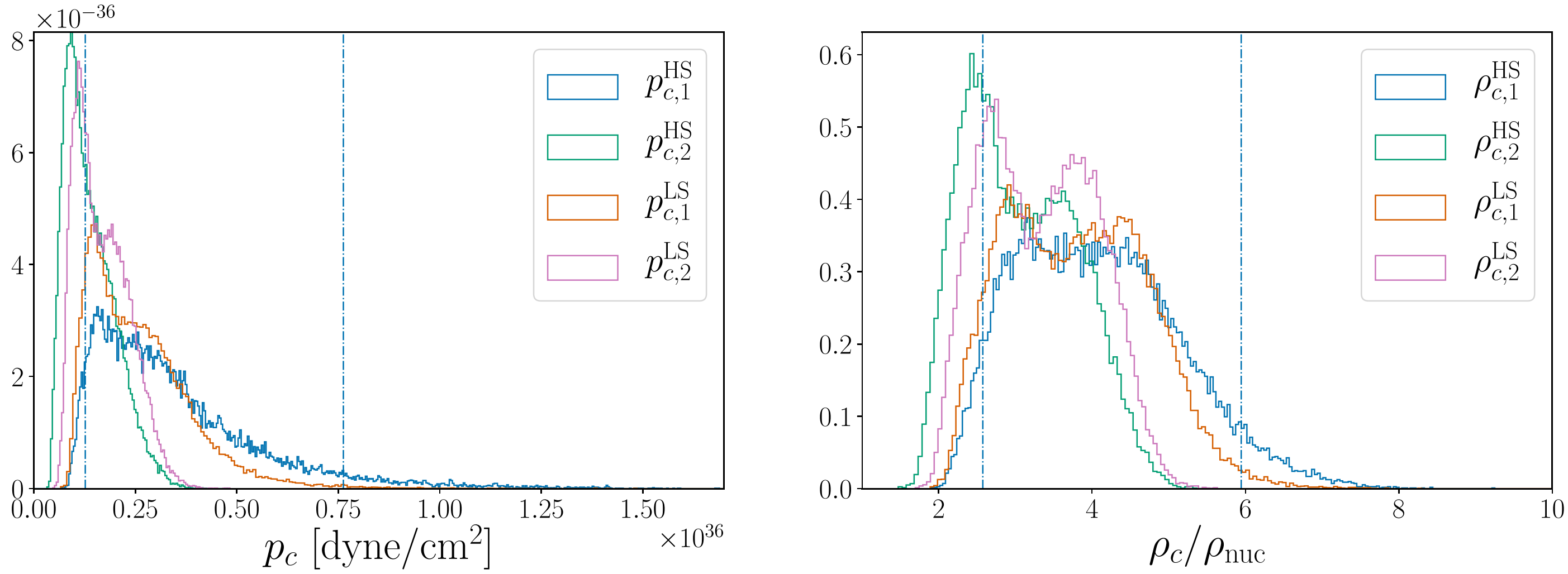

**Figure 17.** Distributions of the pressure (left) and central density (right) obtained with high- and low-spin priors (HS and LS). Vertical dashed lines mark the 90% credible interval. The moderately high primary mass implies central densities ranging from three up to six times nuclear saturation density.

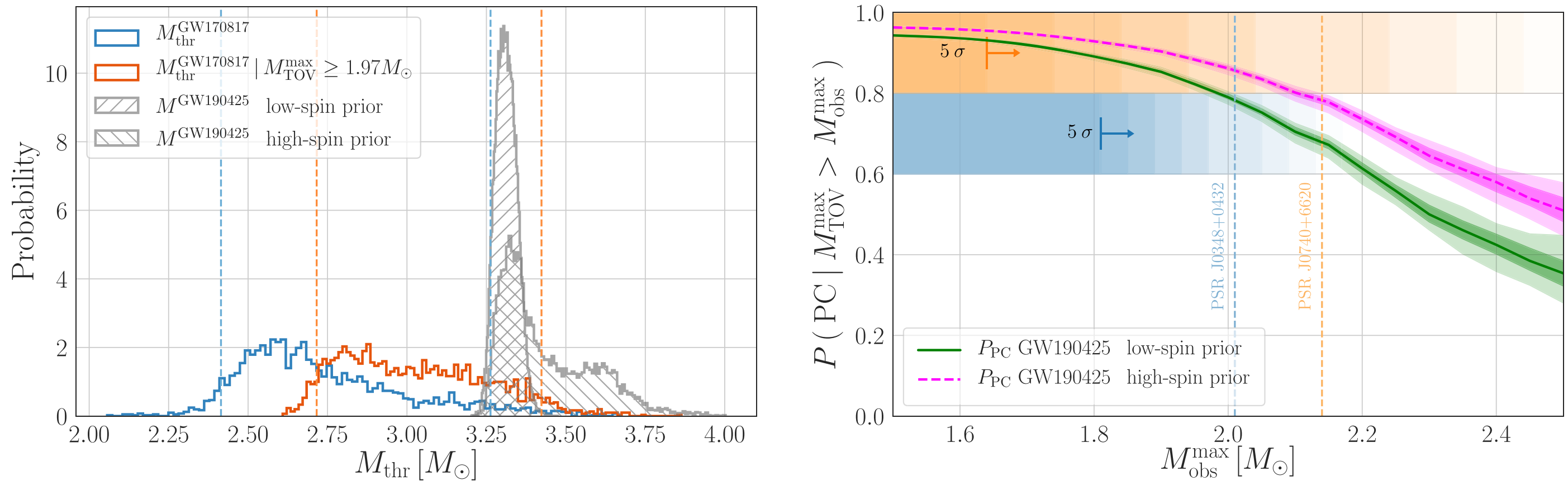

**Figure 18.** Left: distributions of the total mass of GW190425, obtained with high-spin and low-spin priors, compared to the distribution of $M_{\rm thr}$, computed from the posterior samples of Abbott et al. (2017b) and required to either support maximum masses of 1.97 $M_\odot$ (red), or not (blue). Right: probability of prompt collapse for GW190425 as a function of the heaviest observed NS mass. The 50% and 90% credible levels are computed by marginalizing over uncertainties on the fit coefficients. Such error estimates are treated as the standard deviations of a bivariate normal distribution. Shaded bands represent exclusion regions of the maximum supported NS mass, based on mass measurements of the heavy pulsars PSR J0348 + 0432 and PSR J0740+6620 approximated as Gaussians, in half-$\sigma$ steps of confidence out to 5$\sigma$; median values are shown as vertical dashed lines.

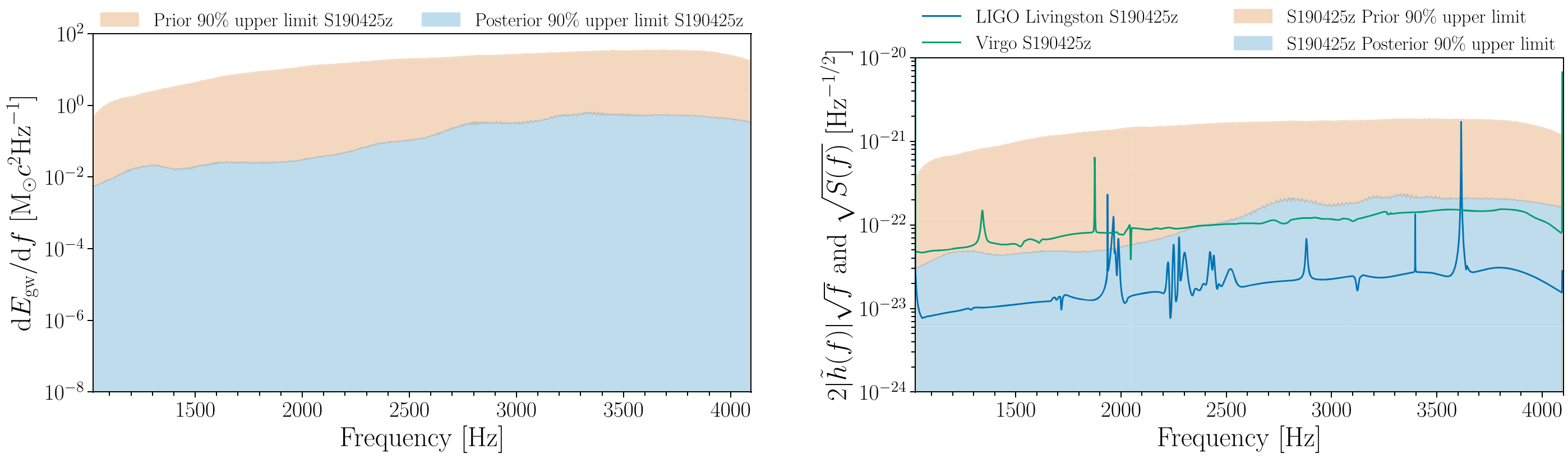

**Figure 19.** 90% credible upper limits on radiated energy (left) and gravitational wave strain (right), and the relative priors. The noise amplitude spectral densities of the Livingston and Virgo instruments used during parameter estimation are also shown for comparison.